\documentclass{jaa}
\usepackage{fix-cm}
\usepackage[T1]{fontenc}
\usepackage{ae,aecompl}
\usepackage[normalem]{ulem}
\usepackage{graphicx}	% Including figure files
\usepackage{amsmath, mathtools, xfrac}	% Advanced maths commands
\usepackage{amssymb, enumitem}	% Extra maths symbols
\usepackage{longtable, multicol}
\usepackage{rotate, setspace}
\usepackage{lscape}
\usepackage{natbib, hyperref}
\usepackage{latexsym,verbatim,xcolor}
\newcommand{\apj}[1]{ApJ, }
\newcommand{\jcp}[1]{J. Chem. Phys., }
\newcommand{\mnras}[1]{MNRAS, }
\newcommand{\aj}[1]{AJ, }
\newcommand{\apjs}[1]{ApJS, }
\newcommand{\apjl}[1]{ApJ Letter, }
\newcommand{\aap}[1]{A\&A, }
\newcommand{\aaps}[1]{A\&A Suppl. Series, }
\newcommand{\araa}[1]{Annu. Rev. A\&A, }
\newcommand{\aaas}[1]{A\&AS, }
\newcommand{\apss}[1]{Ap\&SS }
\newcommand{\bain}[1]{Bul. of the Astron. Inst. of the Netherlands,}
\newcommand{\planss}[1]{Planetary and Space Science,}
\newcommand{\nat}[1]{Nature,}
\newcommand{\actaa}[1]{Acta Astronomica,}
\newcommand{\aapr}[1]{The Astronomy and Astrophysics Review,}
\newcommand{\memsai}[1]{Memorie della Societa Astronomica Italiana,}
\newcommand{\physscr}[1]{Physica Scripta}
\begin{document}\sloppy
%\color{red}
%%paper title
%%For line breaks \\ can be used within title
\title{An In-depth Analysis of Photometric and Kinematic Characteristics of SAI 16, SAI 81 and SAI 86 Open Clusters Utilizing \textit{Gaia} DR3}

\author{A. Y. Alzhrani\textsuperscript{1}, A. A. Haroon\textsuperscript{1}, W. H. Elsanhoury\textsuperscript{2,*}, and D. C. \c{C}{\i}nar\textsuperscript{3}}
\affilOne{\textsuperscript{1}Astronomy and Space Science Department, Faculty of Science, King Abdulaziz University, Jeddah, Saudi Arabia.\\}
\affilTwo{\textsuperscript{2}Physics Department, College of Science, Northern Border University, Arar, Saudi Arabia.\\}
\affilThree{\textsuperscript{3}Programme of Astronomy and Space Sciences, Institute of Graduate Studies in Science, Istanbul University, 34116, Beyaz{\i}t, Istanbul, Turkey.\\}

\twocolumn[{

\maketitle

%%include \corres to print the corresponding author Email id
\corres{elsanhoury@nbu.edu.sa}

%%include \msinfo for
%%manuscript information such as
%%received, revised and accepted dates
%%
\msinfo{22 JAN 2025}{21 APR 2025}

%%abstract
\begin{abstract}
This study comprehensively analyzes three open star clusters: SAI 16, SAI 81, and SAI 86 using \textit{Gaia} DR3 data. Based on the ASteCA code, we determined the most probable star candidates ($P\geq50\%$) and estimated the number of star members of each cluster as 125, 158, and 138, respectively. We estimated the internal structural parameters by fitting the king model to the observed RDPs, including the core, limited, and tidal radii. The isochrone fitting to the color-magnitude diagram provided $\log$ age of 9.13 $\pm$ 0.04, 8.10 $\pm$ 0.04, and 8.65 $\pm$ 0.04 and distances ($d$) of 3790 $\pm$ 94 pc, 3900 $\pm$ 200 pc, and 3120 $\pm$ 30 pc for SAI 16, SAI 81, and SAI 86, respectively. As well, we have calculated their projected distances from the Galactic plane ($X_{\odot}$, $ Y_{\odot})$ as well as their projected distance from the Galactic plane $(Z_{\odot})$, the distance from the Galactic center ($R_{\rm gc}$), and the total mass ($M_{\rm C}$) in solar units are about $142\pm12$, $302\pm17$, and $192\pm14$ for SAI 16, SAI 81, and SAI 86, respectively. Examining the dynamical relaxation state indicates that all three clusters are dynamically relaxed. By undertaking a kinematic analysis of the cluster data, the space velocity was determined. We calculated the coordinates of the apex point  ($A_o,~D_o$) using the AD diagram method along with the derivation of the solar elements ($S_{\odot}$, $l_A$, $b_A$). Through our detailed dynamic orbit analysis, we determined that the three SAI clusters belong to the young stellar disc, confirming their membership within this component of the Galactic structure.

\end{abstract}

%%insert keywords separated by 3 hyphens using \keywords{words}
\keywords{Open star clusters: SAI clusters- Color-Magnitude diagrams- ASteCA Package- Initial mass function-Kinematics.}
}]
%%close the twocolumn escape here

%%include \doinum{number}for the DOI number in the header
%%include \volnum{number} for the volume number in the header
%%include \year{yyyy} for  year of publication in the header
%%include \pgrange{num--num} page range of article in the header
%%include \artcitid{num} for the article citation id
%%include \lp to print last page of the article
%%include \setcounter{page}{pagenum} for the exact starting page of the article

\doinum{12.3456/s78910-011-012-3}
\artcitid{\#\#\#\#}
\volnum{000}
\year{0000}
\pgrange{1--}
\setcounter{page}{1}
\lp{1}

\section{Introduction}
\label{sec1}

As fundamental astrophysical objects, stars are essential in advancing our understanding of the cosmic structure and dynamics, serving as key indicators for unraveling the underlying mechanisms of the universe. Delving into their formation processes stands among the most captivating and essential pursuits in contemporary astrophysics. Notably, stars rarely form in isolation; instead, they are typically born within stellar clusters, which are aggregates of stars sharing a common origin. These clusters emerge from vast reservoirs of gas and dust, known as Giant Molecular Clouds (GMCs), and maintain gravitational binding. The stars within such clusters exhibit a remarkable uniformity in attributes like age, distance, spatial motion, and chemical composition. At the same time, their masses constitute the primary distinguishing factor, driving diverse evolutionary trajectories \citep{LadaandLada2003}.

Among the various types of stellar groupings, open star clusters (OCs) occupy a significant place. These clusters encompass star populations ranging from a few dozen to several hundred, predominantly residing within the Galactic disk \citep{Cantat-Gaudin2018}. Their physical and chemical homogeneity makes OCs invaluable assets in the study of stellar and galactic astrophysics. OCs serve as dynamic astrophysical laboratories, enabling in-depth investigations into stellar formation, evolution, and interstellar interactions. Their utility extends to large-scale galactic studies, offering insights into processes such as the dynamics of the Galactic disk and the structural evolution of the Milky Way (MW). For instance, \citet{Sandage1957} and \citet{Koester1996} highlight the role of OCs in testing theoretical models of stellar formation and assessing stellar interactions within dense environments. Furthermore, the chemical and kinematic properties of OCs have proven instrumental in tracing Galactic abundance gradients and informing our understanding of chemical evolution across the Galaxy \citep{Chen2003, Bonatto2006, Castro-Ginard2021}.

Numerous studies emphasize the significance of OCs in understanding Galactic kinematics. For example, \citet{Dias2005} utilized kinematic data from OCs to reconstruct their orbital histories, thereby identifying their birthplaces and contributing to our knowledge of the rotational dynamics of spiral arms. Similarly, \citet{Hao2021} employed kinematic analyses of OCs to investigate the structure and evolutionary mechanisms governing the MW's spiral arm configuration. Beyond their kinematic contributions, OCs offer valuable data for exploring the Galactic abundance gradient. Research efforts such as those by \citet{Magrini2009} and \citet{Netopil2022} underscore the role of OCs in elucidating the chemical enrichment processes that shape the Galactic disk. 

The spatial distribution of OCs relative to the Galactic center helps us understand the structure and dynamics of the MW. By examining OCs at varying distances from the Galactic center, researchers can delve deeper into the characteristics of the galaxy's stellar populations and the mechanisms influencing its evolutionary pathways \citep{Monteiro2021}. In addition, OCs serve as essential benchmarks for calibrating the astronomical distance scale, as highlighted by \citet{Friel2013}.

The revolutionary contributions of the \textit{Gaia} mission have dramatically enhanced the study of OCs, providing a wealth of high-precision astronomical and photometric data. These unparalleled measurements of stellar positions, motions, and parallaxes have redefined membership criteria and improved the determination of fundamental parameters for numerous OCs \citep{Prusti2016, Cantat-Gaudin2018, Dias2021, Tarricq2022}. The \textit{Gaia} mission has also facilitated detailed analyses of individual cluster properties \citep{Monteiro2020, Akbulut2021, Sariya2021, Badawy2023} and enabled the discovery of many previously unknown OCs \citep{Sim2019, Ferreira2021, Castro2022, Hao2022}.

This paper is structured as follows: Section \ref{sec2} details the datasets employed in this study. In Section \ref{sec3}, we present the results of our analysis using \textit{Gaia} DR3 data, focusing on the astrometric, photometric, dynamical, kinematic, and orbital parameters of selected open clusters. Section \ref{sec4} devoted to discussion of the obtained results. In conclusion, Section 5 summarises the key findings and their implications.

% Table 1
\begin{table*}
\setlength{\tabcolsep}{6pt}
\renewcommand{\arraystretch}{0.9}
\footnotesize
\centering
\caption{Astrophysical parameters for the SAI 16, SAI 81, and SAI 86 OCs are listed in the table, including V-band extinction ($A_{\rm V}$), trigonometric parallax ($\varpi$), iron abundance ([Fe/H]), age ($\log(t)$), proper motion components $(\mu_\alpha\cos \delta,~\mu_\delta)$, and radial velocity ($V_{\rm R}$). Corresponding references are provided in the 'Ref' column.}
\begin{tabular}{cccccccc}
\hline
$A_{\rm V}$ & $\varpi$ & [Fe/H] & $\log(t)$ &  $\langle\mu_{\alpha}\cos\delta\rangle$ &  $\langle\mu_{\delta}\rangle$ & $V_{\rm R}$ & Ref. \\
(mag) &  (mas) & (dex) &  & (mas yr$^{-1}$) & (mas yr$^{-1}$) & (km s$^{-1})$ &      \\
\hline
\hline
\multicolumn{8}{c}{SAI\,16}\\
\hline
--- & 0.157 $\pm$ 0.009 & --- & ---& --- & --- & -65.59 $\pm$ 0.40 &  (1) \\
--- &  --- & -0.19 $\pm$ 0.01 &  --- & --- & --- & -66.10 $\pm$ ~0.80 & (2) \\
--- &  0.161  $\pm$ 0.065 & -0.191 $\pm$ 0.01 & 9.00 & -0.741$\pm$ 0.071& -0.344 $\pm$ 0.087 & --- & (3) \\
2.68 & 0.157 $\pm$ 0.065 & ---& 9.00  & -0.725 $\pm$ 0.305  & -0.305 $\pm$ 0.121 & --- &  (5) \\
--- & 0.209	$\pm$ 0.054 & ---& 9.00 & -0.793 $\pm$ 0.029 & -0.404 $\pm$ 0.021 & --- & (6) \\
--- &  --- & -0.28 $\pm$ 0.01 & 9.00 & --- & --- & -61.20 $\pm$ ~0.20 & (8) \\
2.54 $\pm$ 0.23 & 0.212 $\pm$ 0.060 & --- & 8.43 $\pm$ 0.21 & -0.799 $\pm$ 0.093 & -0.349 $\pm$ 0.093 & -56.26 $\pm$ 13.42 & (11) \\
\hline
\hline
\multicolumn{8}{c}{SAI\,81}\\
\hline
--- & 0.276 & --- & 8.13 $\pm$ 0.08 & -1.991 & 2.931 & --- & (4)  \\
1.53 & 0.209 $\pm$ 0.033 & --- & 8.21 & -2.003 $\pm$ 0.059 & 2.944 $\pm$ 0.059 & --- &  (5)  \\
--- & 0.227	$\pm$ 0.023 & --- & 8.21 & -1.996 $\pm$ 0.028 & 2.965 $\pm$ 0.053 & --- &   (6) \\
1.81  $\pm$ 0.25 &0.214 $\pm$ 0.035 & -0.07 $\pm$ 0.117 & 8.37 $\pm$ 0.59 & -2.001 $\pm$ 0.064 & 2.948 $\pm$ 0.070 & 60.68 $\pm$ 8.32 & (7)\\
1.65 $\pm$ 0.18 & 0.224 $\pm$ 0.033 & --- & 8.28 $\pm$ 0.19 & -2.001 $\pm$ 0.057 & 2.956 $\pm$ 0.073 & 54.97 $\pm$ 0.95 & (11) \\
\hline
\hline
\multicolumn{8}{c}{SAI\,86}\\
\hline
--- & --- & --- & ---& --- & --- & 60.97 $\pm$ 0.98 &  (1) \\
1.72 & 0.219 $\pm$ 0.050 & --- & 8.75 & -3.111 $\pm$ 0.118 & 2.931	$\pm$ 0.116 & --- &  (5) \\
--- & 0.239 $\pm$ 0.043  & --- & 8.75 & -3.107  $\pm$ 0.046 & 2.982  $\pm$ 0.052 & --- &  (6) \\
1.99 $\pm$ 0.12 & 0.211 $\pm$ 0.054 & -0.11 $\pm$ 0.13 & 8.85 $\pm$ 0.13 & -3.124 $\pm$ 0.124 & 2.949 $\pm$ 0.148 & --- & (7) \\
--- & 0.274 $\pm$ 0.030 & --- & 8.65 & -3.017 $\pm$ 0.032 & 3.210 $\pm$ 0.037 & 65.90 $\pm$ 2.00 &  (9) \\
1.72 $\pm$ 0.11 & 0.249 $\pm$ 0.023 & -0.03 $\pm$ 0.16 & 9.01 $\pm$ 0.08 & -- & --- & --- &(10) \\
--- & 0.249 $\pm$ 0.038 & --- & 8.54 $\pm$ 0.17 & -3.083 $\pm$ 0.085 & 2.984 $\pm$ 0.092 & 60.98 $\pm$ 2.47 & (11) \\
\hline
\end{tabular}%
\\
\raggedright
\vspace{5pt}
{\scriptsize (1)~\citet{Soubiran18}, (2)~\citet{2019A&A...623A..80C},(3)~\citet{2021MNRAS.503.3279S}, (4)~\citet{Bouma2019}, (5)~\citet{CantatGaudin2020}, (6)~\citet{Poggio2021}, (7)~\citet{Dias2021}, (8)~\citet{Netopil2022}, (9)~\citet{Liu2023}, (10)~\citet{Cavallo2024}, (11)~\citet{Hunt2024}},
\label{Tab: 1}%
\end{table*}%

%===================================================================
\section{Data}\label{sec2}

The third data release of the \textit{Gaia} mission, referred to as \textit{Gaia} DR3 \citep{GaiaDR3}, serves as the primary dataset for the analysis of the three OCs under investigation. Released on June 13, 2022, this dataset builds upon the \textit{Gaia} DR3 and incorporates a wealth of new astrometric and photometric information. Specifically, it provides multi-band photometry across three bands: $G$, $G_\text{BP}$, and $G_\text{RP}$, alongside radial velocity measurements for more than 33 million celestial sources. The astrometric precision of \textit{Gaia} DR3 has seen a notable enhancement compared to previous releases, with parallax uncertainties ranging from 0.02 to 0.03 mas for brighter stars ($G < 15$ mag) and extending up to 1.3 mas at $G = 21$ mag. Likewise, the uncertainties in proper motion are similarly reduced, with values of 0.02 to 0.03 mas yr$^{-1}$ for $G < 15$ mag stars, and 1.4 mas yr$^{-1}$ for $G = 21$ mag.

To study the clusters, data were extracted from \textit{Gaia} DR3 within a 20 arcmin radius from the centers of the clusters. These data include the three-band photometric measurements $(G,~G_{\rm BP},~G_{\rm RP})$ as well as key astrometric parameters, namely the equatorial coordinates ($\alpha$,~$\delta$), proper motion components $(\mu_\alpha\cos \delta,~\mu_\delta)$ and trigonometric parallax ($\varpi$), along with their respective uncertainties.

\begin{figure}
\centering
\caption{DSS images for the OCs under study. The red plus sign indicates the cluster center, while the red circle marks the cluster region. The density contour diagrams illustrating the structural properties of the SAI 16, SAI 81, and SAI 86 OCs are presented.}
\includegraphics[width=0.99\linewidth]{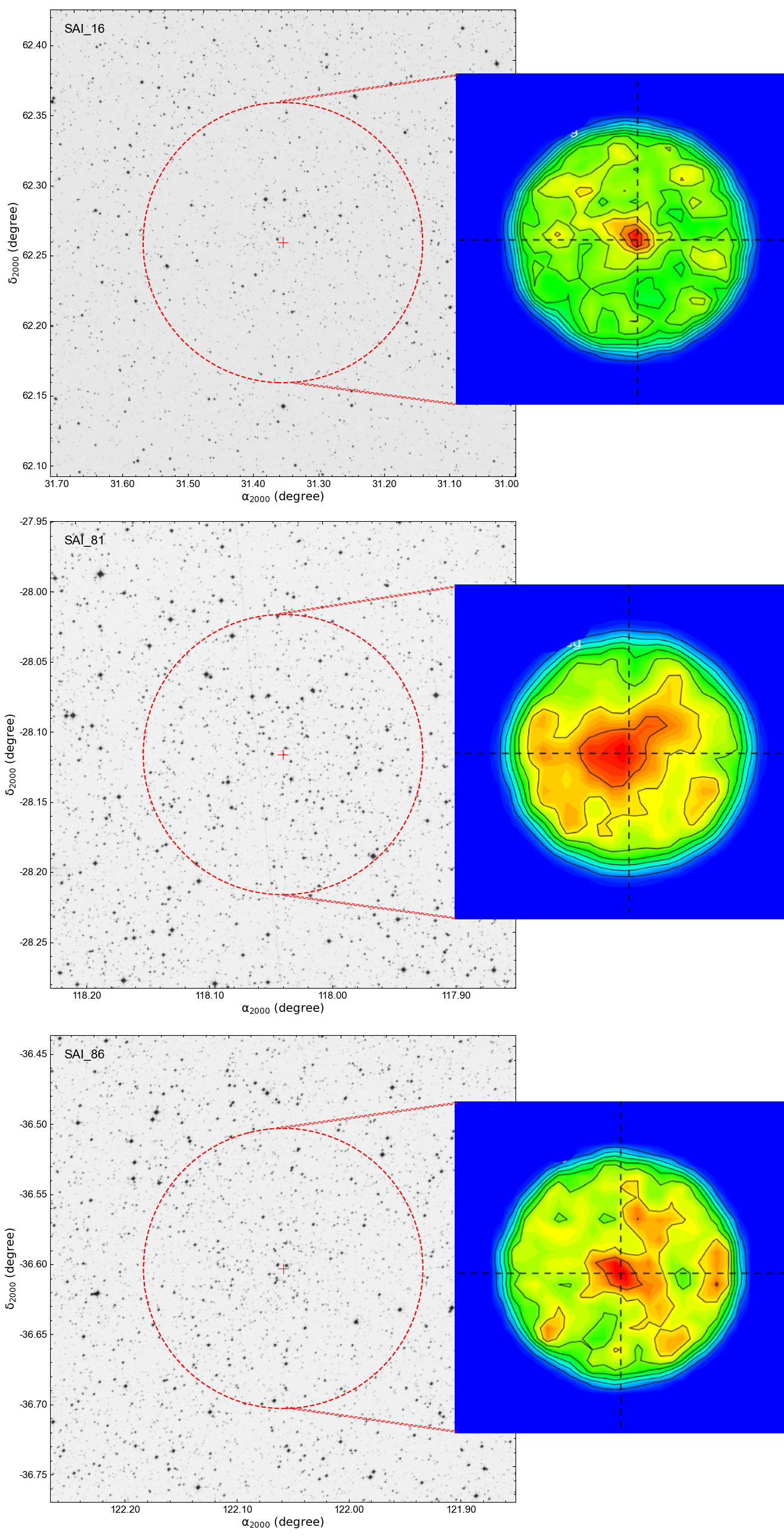}
\label{Fig: 1}
\end{figure}

In this analysis, we focus on three OCs: SAI 16, SAI 81, and SAI 86, selected from a catalog compiled by a research group at the Sternberg Astronomical Institute (SAI)\footnote{\url{http://ocl.sai.msu.ru/catalog/}} in Russia \citep{koposov2008, Glushkova2010}. This catalog provides the initial parameters of OCs identified using data from the Two Micron All Sky Survey (2MASS) \citep{Skrutskie2006}. While numerous studies have investigated various clusters within this catalog, contributing significantly to the field of open cluster research (e.g., \cite{Yadav2014, Carraro2017, Elsanhoury2019, Maurya2021}), these three specific clusters remain relatively unexplored both photometrically and dynamically. Our work presents a comprehensive analysis based on accurate membership assignment using \textit{Gaia} DR3 data for these selected OCs. The initial parameters for these clusters, as listed in the catalog, are provided in Table \ref{Tab: 1}, while the corresponding images from the Digitized Sky Survey (DSS) images are shown in Figure \ref{Fig: 1}.

The fundamental properties of the OCs analyzed in this study were obtained from the SAI Open Cluster Catalog. The equatorial coordinates (J2000) of SAI 16, SAI 81, and SAI 86 are $(\alpha,~\delta) = (02^{\text{h}}05^{\text{m}}30^{\text{s}}, +62^{\circ}15'54'')$, $(07^{\text{h}}52^{\text{m}}07^{\text{s}}, -28^{\circ}07'21'')$, and $(08^{\text{h}}08^{\text{m}}15^{\text{s}}, -36^{\circ}36'33'')$, respectively. Their corresponding Galactic coordinates are $(l,~b) = (131^{\circ}.430, 0^{\circ}.621)$ for SAI 16, $(244^{\circ}.584, -0^{\circ}.546)$ for SAI 81, and $(253^{\circ}.593,~-2^{\circ}.117)$ for SAI 86. The integrated $(B-V)$ colors of these clusters are reported as $0.70 \pm 0.01$ mag for SAI 16, $0.40 \pm 0.02$ mag for SAI 81, and $0.69 \pm 0.09$ mag for SAI 86. The ages ($\log \text{age}$ in years) are determined as $9.15 \pm 0.05$ for SAI 16, younger than $8.70$ for SAI 81, and $8.60 \pm 0.05$ for SAI 86.

%===================================================================
\section{Analysis and Results}\label{sec3}

In this study, we utilized the Automated Stellar Cluster Analysis (ASteCA) code \citep{Perren2015} to conduct a detailed examination of the stellar clusters under consideration. ASteCA\footnote{\url{http://asteca.github.io/}} is an advanced Python-based software package specifically developed for the automated analysis and characterization of stellar clusters. It is widely adopted in astronomical research for the systematic investigation of stellar populations within open and globular star clusters (SCs). The code enables the determination of structural and astrophysical parameters of SCs, age ($t$), metallicity ([Fe/H]), distance $d$, $V$-band extinction ($A_{\rm V}$), and various other intrinsic properties. By employing sophisticated statistical techniques, ASteCA streamlines the process of analyzing photometric data from stellar surveys to derive these fundamental cluster characteristics. This automation not only accelerates the analysis but also enhances the precision and consistency of the derived results \citep{Perren2015}.

\subsection{Structural Parameters}
%======== Center and radius assignment

Accurately determining the central coordinates of a stellar cluster represents a fundamental step in characterizing its physical and dynamical properties. In this study, the identification of the cluster centers was performed using the ASteCA code, which employs a two-dimensional Gaussian kernel density estimator (KDE) to analyze the spatial distribution of stars within the cluster. This method systematically identifies the region of the highest stellar density within the observational frame, which is then designated as the cluster center. An illustrative example of this procedure is provided in Figure \ref{Fig: 1}.

The re-calibrated central coordinates were found to exhibit strong concordance with the values reported by \cite{Glushkova2010} along the right ascension axis, while slight deviations were observed in the declination direction. This minor adjustment underscores the precision of the KDE approach in refining the cluster center positions. Subsequently, the updated equatorial $(\alpha,~\delta)$ and Galactic $(l,~b)$ coordinates for each cluster are presented in Table \ref{Tab: 2}, providing a robust foundation for further analysis of their structural and kinematic properties.

\begin{table*}
\caption{The newly determined central coordinates of the three clusters are presented in both equatorial ($\alpha$, $\delta$) and Galactic ($l$, $b$) coordinate systems.}
\centering
\begin{tabular}{lcccc}
\hline
\textbf{Cluster} & \textbf{$\alpha$} & \textbf{$\delta$} & \textbf{$l$} & \textbf{$b$} \\
\hline
\hline
SAI 16 & $2^{\text{h}}\,5^{\text{m}}\,31^{\text{s}}.50$ & $+62^{\circ}\,15'\,55''.90$ & $131^{\circ}$.4340& $+0^{\circ}$.6230 \\
SAI 81 & $7^{\text{h}}\,52^{\text{m}}\,03^{\text{s}}.85$ & $-28^{\circ}\,09'\,27''.65$ & $244^{\circ}.6084$ & $-0^{\circ}.5742$ \\
SAI 86 & $8^{\text{h}}\,8^{\text{m}}\,14^{\text{s}}.95$ & $-36^{\circ}\,36'\,57''.85$ & $253^{\circ}.5988$ & $-2^{\circ}.1211$ \\
\hline
\end{tabular}
\label{Tab: 2}
\end{table*}

\begin{figure}
\centering
\includegraphics[width=0.95\linewidth]{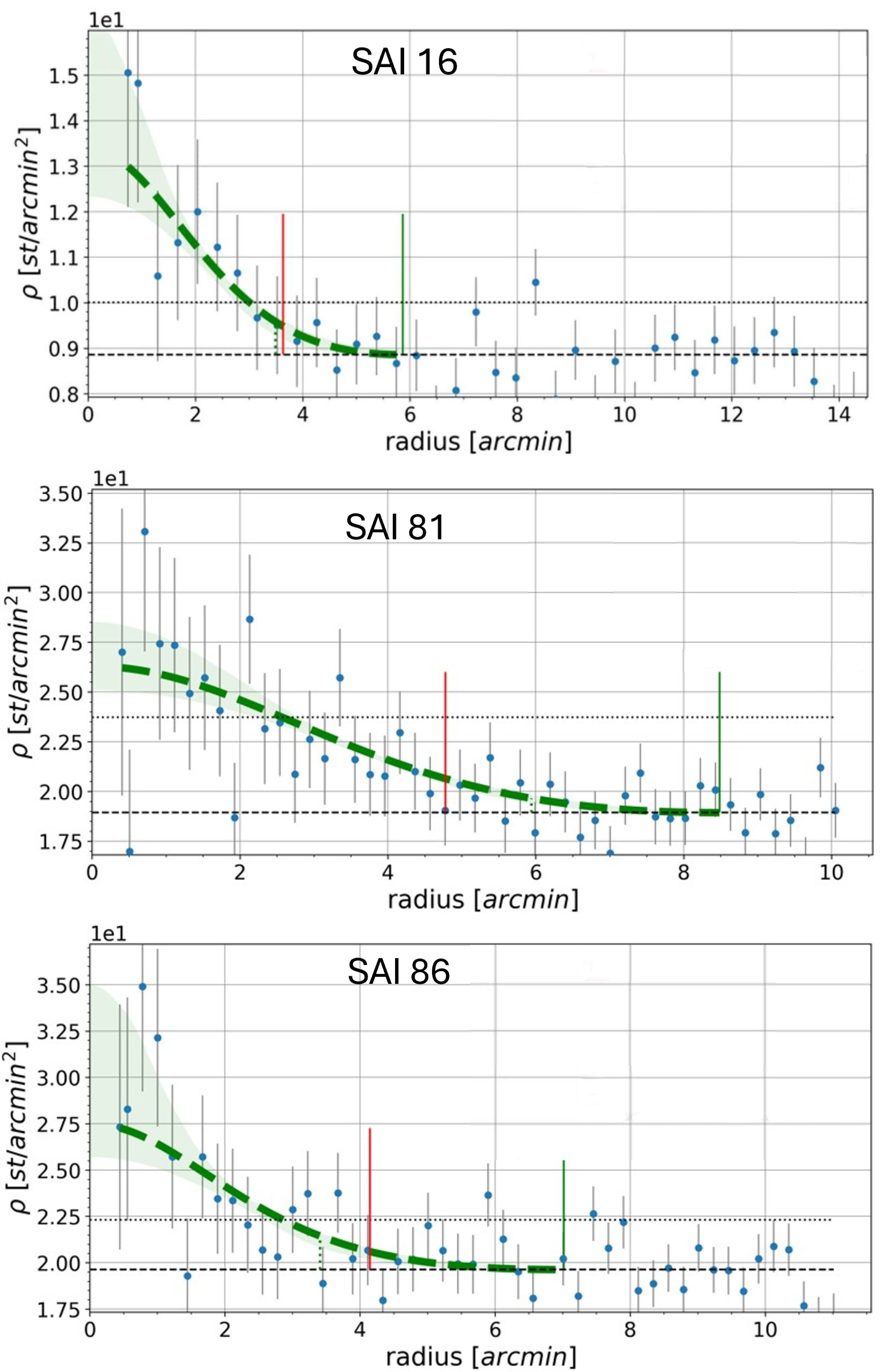}
\caption{King profile fitting to the surface density extracted RDP parameters. Blue dots show the ASteCA RDP, green dashed curve the King profile, black lines $\rho_{\circ}$ and $\rho_{\text{bg}}$, and vertical lines mark $r_{\rm c}$, $r_{\rm cl}$, and $r_{\rm t}$.}
\label{Fig: 3}
\end{figure}

To investigate the surface density distribution of the clusters, it is essential to construct their radial density profiles (RDPs), as depicted in Figure \ref{Fig: 3}. The RDP represents the variation in the stellar surface density as a function of radial distance from the cluster center, effectively illustrating the spatial distribution of stars within the cluster \citep{Haroon2014}. In this study, the RDP was generated using the ASteCA code, which constructs concentric square rings based on an underlying two-dimensional histogram grid in the positional space. For each ring, the stellar surface density ($\rho$) is computed by dividing the number ($N$) of stars within the ring by the corresponding area ($A$) of the ring, expressed as $\rho = N/A$. 

To characterize the observed RDP, it was fitted with a King profile, a widely used analytical model described by \cite{King1962}. The King profile is given by $\rho(r) = \rho_{\rm bg} + \rho_o / [1 + (r/r_{\rm c})^2]$, where $\rho_o$ represents the central surface density, and $\rho_{bg}$ corresponds to the background surface density. The parameter $r_c$, known as the core radius, defines the distance from the center of the cluster at which the stellar density $\rho(r)$ drops to half of its central value $\rho_o$. 

ASteCA also identifies the limiting radius $(r_{\rm cl})$, which is the distance from the center of the clusters at which the RDP merges with the background density level. For the clusters analyzed in this study, the limiting radii were determined to be $3.63$, $4.78$, and $4.15$ arcmin for SAI 16, SAI 81, and SAI 86, respectively.

In this study, we have determined the density contrast parameter ($\delta_{\rm c}$) and the concentration parameter ($C$) for the first time for these clusters, providing a detailed characterization of their structural properties. The density contrast parameter $\delta_{\rm c}$, defined as $\delta_{\rm c} = 1 + \rho_{\rm o}/\rho_{\rm bg}$, quantifies the relative stellar density of the cluster compared to the surrounding background ($\rho_{\rm o}$ and $\rho_{\rm bg}$ represent the central and background densities, respectively). Higher values of $\delta_{\rm c}$, within the range $7 \lesssim \delta_{\rm c} \lesssim 23$, signify the compactness typically associated with dense star clusters \citep{BonattoBica2009}. 

Additionally, the concentration parameter $C$, which offers insights into the distribution of stars within the cluster, is expressed as $C = r_{\rm cl}/r_c$, where $r_{\rm cl}$ and $r_{\rm c}$ are the limiting and core radii, respectively \citep{King1966}. For the clusters analyzed in this study, $C$ values varied between 0.80 and 1.12, reflecting different levels of central condensation and structural compactness. The structural parameters obtained for SAI 16, SAI 81, and SAI 86 such as the core, limiting, and tidal radii, central surface density, background surface density, density contrast, and concentration parameters are comprehensively presented in Table \ref{Tab: 3}.

\begin{table}[t]%
\centering
\footnotesize
\caption{The structural parameters of the clusters under investigation were derived by applying the King profile fitting method to the observed surface density distribution. \label{Tab: 3}.}%
\tabcolsep=0pt%
\renewcommand{\arraystretch}{1.3}
\begin{tabular*}{20pc}{@{\extracolsep\fill}lccc@{\extracolsep\fill}}
\hline
\textbf{Parameters} & \textbf{SAI 16}  & \textbf{SAI 81}  & \textbf{SAI 86}  \\
\hline
\hline
$r_c$ (arcmin) & $3.49^{+4.72}_{-2.25}$   & $5.94^{+7.42}_{-4.49}$  & $3.41^{+4.88}_{-2.11}$   \\
$r_c$ (pc) & $3.87^{+5.23}_{-2.50}$   & $6.80^{+8.49}_{-5.14}$  & $3.63^{+5.19}_{-2.25}$   \\
$r_{\rm cl}$ (arcmin) & 3.63  & 4.78  & 4.15  \\
$r_{\rm cl}$ (pc) & 4.03  & 5.47  & 4.42  \\
$r_{\rm t}$ (arcmin) & $5.87^{+6.78}_{-4.98}$  & $8.49^{+9.23}_{-7.76}$  & $7.02^{+7.92}_{-6.07}$  \\
$r_{\rm t}$ (pc) & $6.51^{+7.52}_{-5.52}$  & $9.72^{+10.56}_{-8.88}$  & $7.47^{+8.43}_{-6.46}$  \\
$ \rho_o$ (stars arcmin$^{-2}$) & 10.31$\pm$3.21  & 28.75$\pm$5.36  & 25.00$\pm$5.00  \\
$\rho_{bg}$ (stars arcmin$^{-2}$) & 8.93$\pm$2.90  & 19.09$\pm$4.37  & 19.74$\pm$4.44    \\
$\delta_{\rm c}$ & 2.16$\pm$0.10&2.51$\pm$0.09&2.27$\pm$0.09\\
$C$  & 1.04  &0.80  & 1.22  \\
\hline
\end{tabular*}
\end{table}

%======== Astrometric parameters and distance determination
\subsection{Astrometric Parameters}

The precise determination of membership probabilities remains one of the most significant challenges in the study of open star clusters. Membership probability quantifies the likelihood that a given star belongs to a cluster rather than being a foreground or background field star that coincidentally aligns along the same line of sight. Accurate determination of cluster membership is paramount, as contamination from field stars can severely distort the derived astrophysical parameters of OCs, such as their age, metallicity, reddening, and distance. Thus, robust membership identification is a foundational step in any rigorous analysis of cluster properties \citep{Carraro2008}. 

Building upon the principle that genuine cluster members exhibit similar kinematic and photometric properties, ASteCA software \citep{Perren2015} implements a Bayesian approach to distinguish true cluster members from field star contaminants. The algorithm systematically compares the distributions of photometric, parallax, and proper motion data of stars within the cluster region against those observed in adjacent field regions. By evaluating thousands of potential scenarios, the algorithm determines the likelihood that each star is a true member of the cluster. Unlike UPMASK \citep{KroneMartins2014}, which is widely used for identifying cluster members in large datasets such as $Gaia$ due to its efficiency in handling large samples, ASteCA provides a more comprehensive analysis by simultaneously determining membership probabilities and estimating key cluster parameters. While UPMASK excels in rapid, large-scale membership assignments without incorporating physical modeling, ASteCA integrates photometric, kinematic, and structural information within a Bayesian framework, allowing for a more detailed characterization of OCs. In some studies, UPMASK is used for an initial membership selection, followed by ASteCA for refined membership determination and parameter estimation. Stars with distributions matching the cluster's properties are assigned higher membership probabilities, while those resembling field stars are assigned lower probabilities \citep{Perren2020}.

In this study, we have adopted a conservative threshold, designating stars with membership probabilities $(P \geq 50\%)$ as likely cluster members. This threshold ensures a balance between including genuine members and minimizing contamination from non-member stars. The application of the ASteCA algorithm to the studied clusters, SAI 16, SAI 81, and SAI 86 yields a highly probable member count of 125, 158, and 138 stars, respectively. These results underscore the efficacy of Bayesian methodologies in refining cluster membership and enhancing the reliability of subsequent astrophysical analyses.

\begin{figure*}
\centering
\includegraphics[width=0.85\linewidth]{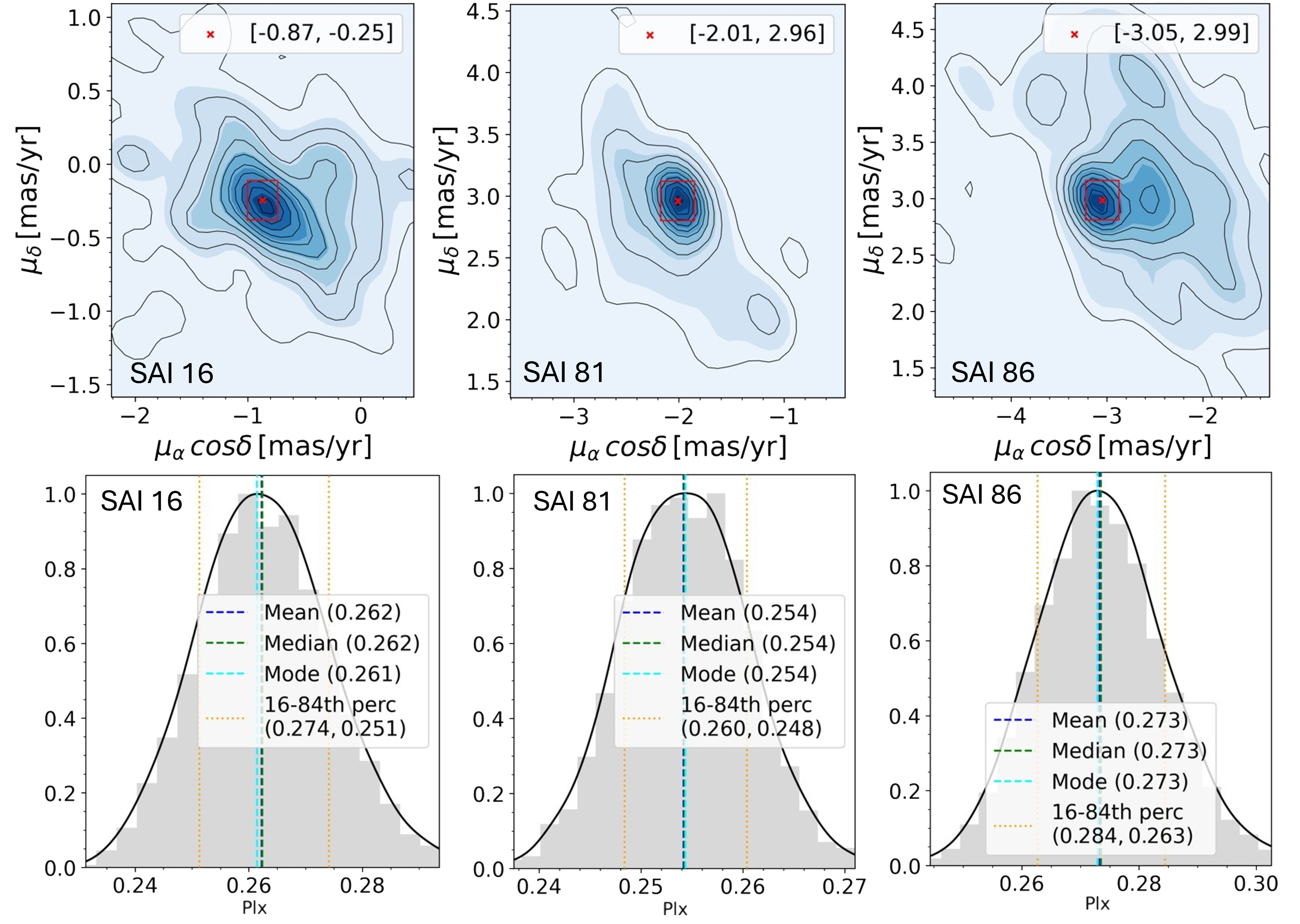}
\caption{Upper panel: The distribution of the mean proper motion dispersions in the right ascension ($\mu_\alpha\cos{\delta}$) and declination ($\mu_\delta$) directions for the clusters. Lower panel: The normalized parallax distribution of potential stellar members within the cluster boundaries.}
\label{Fig: 4}
\end{figure*}

To investigate the kinematic properties of the clusters, the stellar distribution in the proper motion space $(\mu_\alpha\cos\delta,~\mu_\delta)$ was analyzed. The upper panels of Figure~\ref{Fig: 4} illustrate these distributions for each cluster, where the clustering of stars in this space highlights the coherence of their motion. To determine the mean proper motion of the clusters, a Gaussian distribution was fitted along both proper motion components, providing robust estimates for $\mu_\alpha\cos\delta$ and $\mu_\delta$. 

In addition to the proper motion analysis, the parallax distributions of the cluster candidate members were examined, as shown in the lower panels of Figure~\ref{Fig: 4}. The histograms depict the parallax values of the identified members, with a Gaussian fit overlaid (black line) to model the distribution. Trigonometric parallax remains the most precise method for measuring stellar distances; however, systematic zero-point errors in astrometric data introduce significant uncertainties, especially for distant clusters. Recent studies \citep[e.g.,][]{Huang2021, Zinn2021, Wang2022, Dursun2024} have addressed this issue by proposing corrections based on extensive trigonometric datasets from {\it Gaia} EDR3/DR3 \citep{GaiaEDR3, GaiaDR3}. Given the distances of SAI 16, SAI 81, and SAI 86, we applied a zero-point correction to the parallaxes of probable cluster members $(P \geq 50\%)$ adopting
$\varpi_{\rm ZP}=-0.025$ mas from \citet{Lindegren2021} and correcting each star’s parallax using $\varpi_0 = \varpi - \varpi_{\rm ZP}$. This fitting allowed for precise determination of the mean trigonometric parallax for each cluster. Leveraging the mean trigonometric parallax values, the distances to the clusters were calculated using the relationship $d_{\varpi} = 1000/\varpi$. The calculated distances ($d_{\varpi}$; pc) are as follows: 3813, 3934, and 3659 for SAI 16, SAI 81, and SAI 86, respectively. Subsequently, the mean trigonometric parallax values and mean proper motion components were obtained for three OCs and listed in Table \ref{Tab: 4}.

\subsection{Astrophysical Parameters}

\begin{figure*}
\centering
\includegraphics[width=0.95\linewidth]{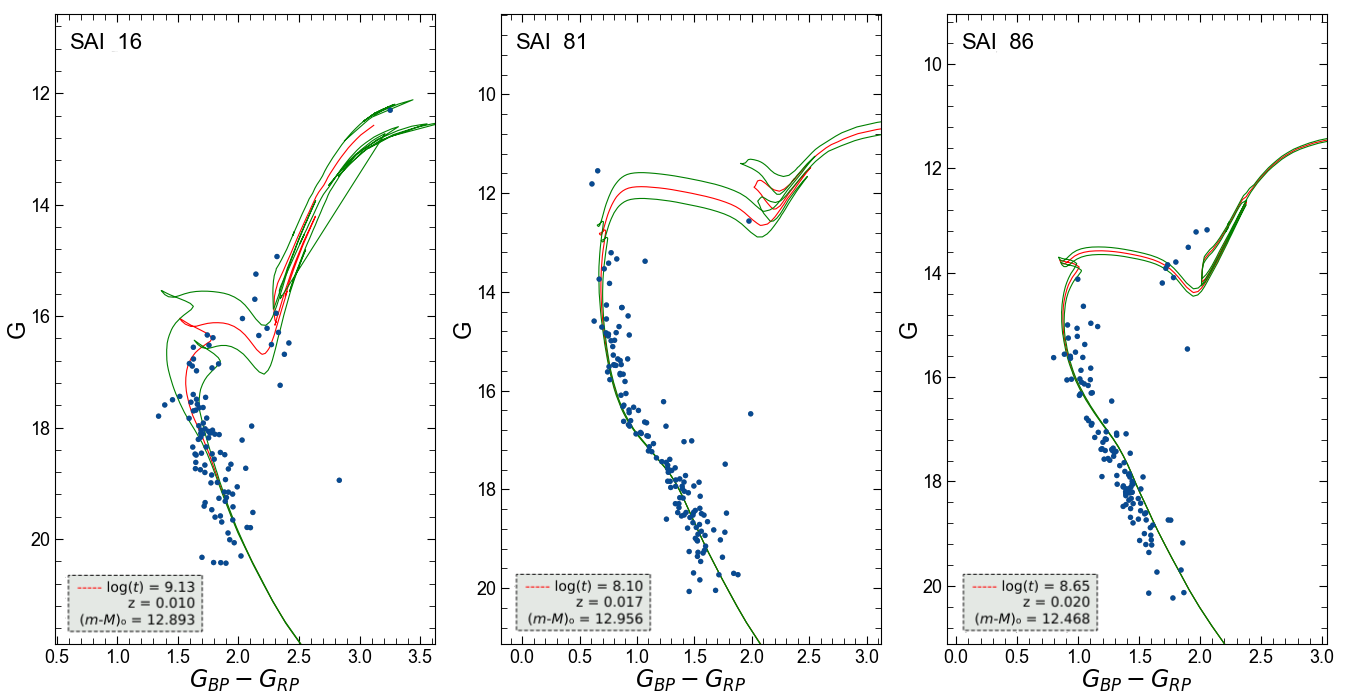}
\caption{In the CMDs for the stellar members of the three open clusters, the red and green lines correspond to the best-fit isochrones and their associated error margins, respectively.}
\label{Fig: 5}
\end{figure*} 

The color-magnitude diagram (CMD) serves as an indispensable tool for deriving the fundamental parameters of stellar clusters, age, distance modulus, color excess, and metallicity. This analysis relies heavily on the precision of CMDs, which are generated from photometric observations and theoretical isochrone models. In this study, we employed ASteCA, a comprehensive cluster analysis tool, to perform isochrone fitting using synthetic CMDs. ASteCA employs a genetic algorithm to identify the optimal match between theoretical isochrones and the observed photometric data of cluster member stars. For our analysis, we utilized the PARSEC v1.2S isochrone models \citep{Bressan2012}, applying them to the photometric magnitudes ($G$, $G_\text{BP}$, and $G_\text{RP}$) of the cluster members, as illustrated in Figure~\ref{Fig: 5}.

The reddening parameter $E(G_\text{BP} - G_\text{RP})$, was derived using the relationship $E(G_\text{BP} - G_\text{RP}) = 1.289~\times~E(B-V)$, where $E(B-V)$ represents the color excess \citep{Elsanhoury2022}. To correct for interstellar extinction, the extinction coefficient $(A_\text{G})$ was computed as $A_\text{G} = 2.74~\times~E(B-V)$, following the methodology outlined by \cite{Casagrande2018} and \cite{Zhong2019}. The distances derived from CMD fitting were found to be consistent with those obtained through astrometric measurements based on cluster parallaxes (i.e., $d_\varpi$). 

The metallicity of SAI 16, SAI 81, and SAI 86 was determined using the isochrone fitting method, considering relevant studies from the literature. For SAI 81 and SAI 86, photometric metallicity estimates from \citet{Dias2021}, as listed in Table \ref{Tab: 1}, were available and used solely for comparison. In contrast, spectroscopic analyses have yielded multiple metallicity estimates for SAI 16. \citet{2021MNRAS.503.3279S} derived [Fe/H] = -0.191 $\pm$ 0.011 dex based on high-resolution spectroscopy of a single member star, while \citet{2019A&A...623A..80C} reported a metallicity of [Fe/H] = -0.19 dex from the analysis of two stars. Both estimates are given relative to the solar reference. These spectroscopic metallicities were adopted as the OC’s metallicity constraint during the isochrone fitting process. To determine the best-fit isochrone and derive the astrophysical parameters for SAI 16, the adopted metallicity [Fe/H] = -0.19 dex was converted into the mass fraction $z$ using the transformation equations provided by Bovy\footnote{github.com/jobovy/isodist/blob/master/isodist/Isochrone.py}, which apply to the {\sc PARSEC} models \citep{Bressan2012}.

\begin{equation}
z_{\rm x} = 10^{{\rm [Fe/H]}+\log \left(\frac{z_{\odot}}{1-0.248-2.78\times z_{\odot}}\right)}
\end{equation}
\begin{equation}
z = \frac{(z_{\rm x}-0.2485\times z_{\rm x})}{(2.78\times z_{\rm x}+1)}.
\end{equation} 
where, $z_{\rm x}$ represents an intermediate parameter, and $z_{\odot}$, the solar metallicity, is taken as 0.0152 \citep{Bressan2012}. Applying these expressions, the corresponding mass fraction for [Fe/H] = -0.19 dex was determined to be $z$ = 0.010.

For SAI 16, our analysis revealed a solar metallicity value of $0.010 \pm 0.003$. The logarithm of the OC’s $\log$ age was determined to be approximately $9.13$, aligning with results from previous studies. However, the distance modulus estimated in this study, $12.893 \pm 0.054$ mag, is notably lower than the value of $14.42 \pm 0.02$ mag reported by \cite{Glushkova2010}. Similarly, the color excess $E(B-V)$ was found to be $0.965 \pm 0.021$ mag, which is consistent with the findings of \cite{kharchenko2016}, and higher than the $0.70 \pm 0.01$ mag reported by \cite{Glushkova2010}. Furthermore, our analysis suggests a cluster distance of $3790 \pm 94$ pc, which is smaller than previous estimates. These discrepancies underscore the importance of integrating CMD-based analyses with astrometric measurements to refine the fundamental parameters of stellar clusters and improve the accuracy of their characterization \cite{Yontan2022}.

For the SAI 81 cluster, the determined solar metallicity is $0.017 \pm 0.002$ dex, and the logarithmic age value is approximately 8.10. This value is in agreement with prior studies, including those of \cite{kharchenko2016} and \cite{CantatGaudin2020}. However, our estimate of the distance modulus, $12.956 \pm 0.106$ mag, is notably larger than the value reported by \cite{Glushkova2010} of $12.21 \pm 0.19$ mag. The color excess for this cluster, measured at $0.70$ mag, shows good consistency with previous findings \citep{kharchenko2016} and our results. The reddening value is determined to be $0.926 \pm 0.010$ mag. Furthermore, our analysis suggests a smaller cluster distance of $3900 \pm 200$ pc, in contrast to the values reported by \cite{kharchenko2016} and \cite{CantatGaudin2020}, which were slightly larger.

For the SAI 86 cluster, we obtained a solar metallicity value of $0.020 \pm 0.001$ dex. The logarithmic age ($\log t$) is approximately $8.65$, which is in line with previous studies, suggesting the consistency of this value across multiple analyses. The distance modulus and the color excess for SAI 86 were found to be in close agreement with earlier literature. However, the distance calculated in our study, $3120 \pm 30$ pc, aligns with the values of \cite{Glushkova2010} and \cite{kharchenko2016}, while it is lower than the value reported by \cite{CantatGaudin2020}.

To further explore the spatial relationships of these clusters within the Galaxy, the distance to the Galactic center, $R_\text{gc}$, can be computed using the following formula:
\begin{equation} \label{Eq: 3}
R_\text{gc} = \sqrt{R_{\circ}^{2}+(d \cos b)^2- 2 R_{\circ} d \cos b \cos l},
\end{equation}
where $R_{\circ}$, the Sun’s distance from the Galactic center, was adopted to be $8.20 \pm 0.10$ kpc \citep{Bland2019}, $d$ represents the cluster's distance, and $b$ and $l$ denote the Galactic latitude and Galactic longitude, respectively, for clusters. Furthermore, the projected distances in the Galactic plane ($X_\odot,~Y_\odot$) and the distance above or below the Galactic plane ($Z_\odot$), were determined where $X_\odot= d \cos b \cos l, \; Y_\odot = d \cos b \sin l$, and $Z_\odot= d\sin b$ \citep{Elsanhoury2022}.
 
\begin{table*}[htbp]
\centering
\caption{A comparative analysis of the astrophysical and photometric parameters derived for SAI 16, SAI 81, and SAI 86 in this study.}
\label{Tab: 4}
\renewcommand{\arraystretch}{1.2} % Line spacing
\setlength{\tabcolsep}{2pt}
\begin{tabular}{lccc}
\hline
\textbf{Parameters}   & \textbf{SAI 16} & \textbf{SAI 81} & \textbf{SAI 86} \\ 
\hline
\hline
No. of members& 125 & 158 & 138 \\
$\mu_\alpha \cos \delta$ (mas yr$^{-1}$) & -0.87 $\pm$ 0.01 & -2.01 $\pm$ 0.71 & -3.05 $\pm$ 0.01 \\
$\mu_\delta$ (mas yr$^{-1}$) & -0.25 $\pm$ 0.02 & 2.96 $\pm$ 0.01 & 2.99 $\pm$ 0.01 \\
$\varpi$ (mas) & 0.262 $\pm$ 0.005 & 0.254 $\pm$ 0.020 & 0.273 $\pm$ 0.020 \\
$d_{\varpi}$ (pc) & 3813 $\pm$ 62 & 3934 $\pm$ 63 & 3659 $\pm$ 61 \\
Z & 0.010 $\pm$ 0.003 & 0.017 $\pm$ 0.002 & 0.020 $\pm$ 0.001 \\
$\log t$   & 9.13 $\pm$ 0.04 & 8.10 $\pm$ 0.04 & 8.65 $\pm$ 0.04 \\
$E(B-V)$ (mag) & 0.965 $\pm$ 0.021 & 0.718 $\pm$ 0.0008 & 0.707 $\pm$ 0.006 \\
$E(G_\text{BP} - G_\text{RP})$ (mag) & 1.243 $\pm$ 0.027 & 0.926 $\pm$ 0.010 & 0.911 $\pm$ 0.008 \\
$A_{\text{G}}$ (mag) & 2.644 & 1.967 & 1.937 \\
$(m-M)_o$ (mag) & 12.893 $\pm$ 0.054 & 12.956 $\pm$ 0.106 & 12.468 $\pm$ 0.018 \\
$d$ (pc) & 3790 $\pm$ 94 & 3900 $\pm$ 200 & 3120 $\pm$ 30 \\
$R_\text{gc}$ (kpc) & 11.422$\pm$3.38 & 10.482$\pm$3.24 & 9.560$\pm$3.10 \\
$X_\odot$ (kpc) & -2.779$\pm$1.67 & -1.672$\pm$0.77 & -0.880$\pm$0.09 \\
$Y_\odot$ (kpc) & 3.149$\pm$1.78 & -3.523$\pm$1.88 & -2.990$\pm$1.73 \\
$Z_\odot$ (kpc) & 0.046$\pm$0.01 & -0.039$\pm$0.02 & -0.115$\pm$0.03 \\
\hline
\end{tabular}
\end{table*}

\subsection{Luminosity and Mass Function}

\begin{figure*}
\centering
\includegraphics[width=0.95\linewidth]{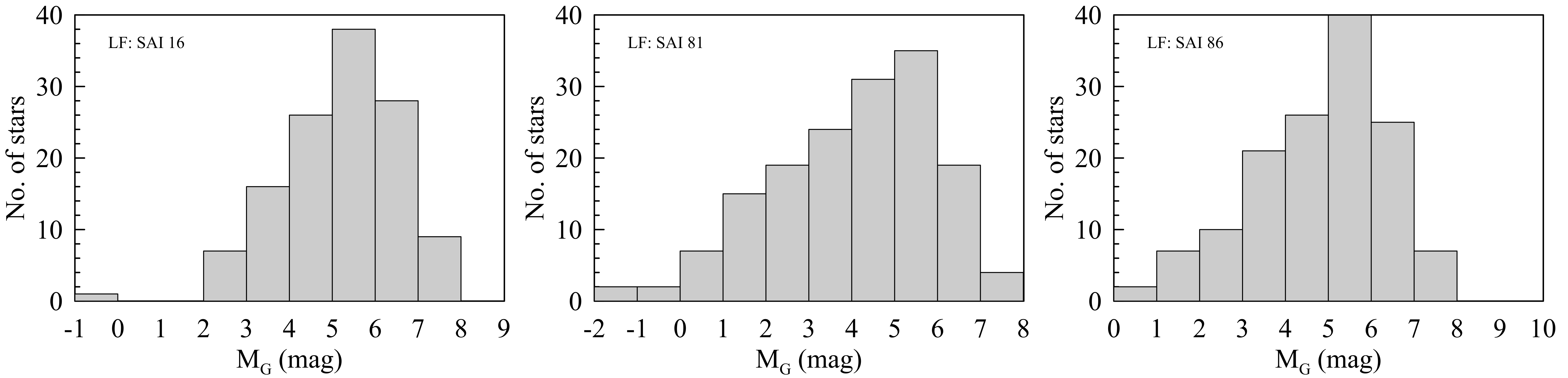}
\caption{The LFs frequency distributions in the $G$ band for the three open clusters analyzed in this study.}
\label{Fig: 6}
\end{figure*}

The Luminosity Function (LF) is an essential tool for understanding the distribution of stellar brightness within a cluster. It quantifies the relative number of stars found within specific intervals of absolute magnitudes \citep{Haroon2017}. As illustrated in Figure \ref{Fig: 6}, the observed LFs in the $G$ band for stars with absolute magnitudes below zero are presented. These LFs are presented in absolute magnitude scales following the application of the distance modulus $(m-M)_o$ derived in the preceding section.

The Initial Mass Function (IMF) provides crucial information on the distribution of stellar masses, serving as a direct consequence of the star formation process \citep{Bisht2019}. Historically, the IMF for stars in the solar neighborhood was first established by \cite{Salpeter1955}, who proposed a power-law model for the mass function (MF). This model describes the number of stars produced within a defined mass range in a given region of space. The MF follows a power law, expressed as:
\begin{equation} \label{Eq: 4}
\dfrac{dN}{dM} \propto M^{-\alpha},
\end{equation}
In this context, $dN/dM$ denotes the number of stars within the mass range $(M, M + dM)$, while $\alpha$ is a constant parameter, conventionally set to 2.35. This value, referred to as the Salpeter slope, signifies that low-mass stars are considerably more abundant than their high-mass counterparts. The stellar masses within the clusters are derived by converting LFs into MFs through the mass-luminosity relation (MLR) based on the selected isochrones with \cite{2018A&A...616A...4E}. To estimate stellar masses within the clusters, we employed a polynomial mass-luminosity relation based on theoretical isochrone fitting:

\begin{equation}
M_{\rm c} = a_0 + a_1 \times M_{\rm G} + a_2 \times M_{\rm G}^2 + a_3 \times M_{\rm G}^3 + a_4 \times M_{\rm G}^4,
\label{Eq: 5}
\end{equation}
where, $a_0, a_1, a_2, a_3$, and $a_4$ are coefficients obtained by fitting the isochrones to the MF of each cluster. These parameters play a crucial role in accurately determining stellar masses, offering key insights into the stellar populations of the clusters. Applying Equation \ref{Eq: 5}, we obtain the estimated total MS mass of about $142\pm12$, $302\pm17$, and $192\pm14$ $M_{\odot}$ for the clusters SAI 16, SAI 81, and SAI 86, respectively, as a function of the age and metalicity such obtained previously. The key parameters characterizing the stellar populations and physical properties of the OCs are summarized in Table \ref{Tab: 5}, which includes the MS total mass ($M_{\text{C}}$) and MS mean mass $\langle M_C \rangle$. These values provide valuable insight into the structure and evolutionary state of the clusters. Significant variations are observed among the clusters, not only in terms of their total mass but also in the number of stars and the mean stellar mass per cluster. These differences reflect the heterogeneity in their stellar populations and the various stages of evolution they may be undergoing. Here, in our calculations, we omitted the post MS stars appear as highlights with CMD of about 18, 3, and 8 stars with respective manner of SAI 16, SAI 81, and SAI 86 OCs.

\begin{figure*}
\centering
\includegraphics[width=0.95\linewidth]{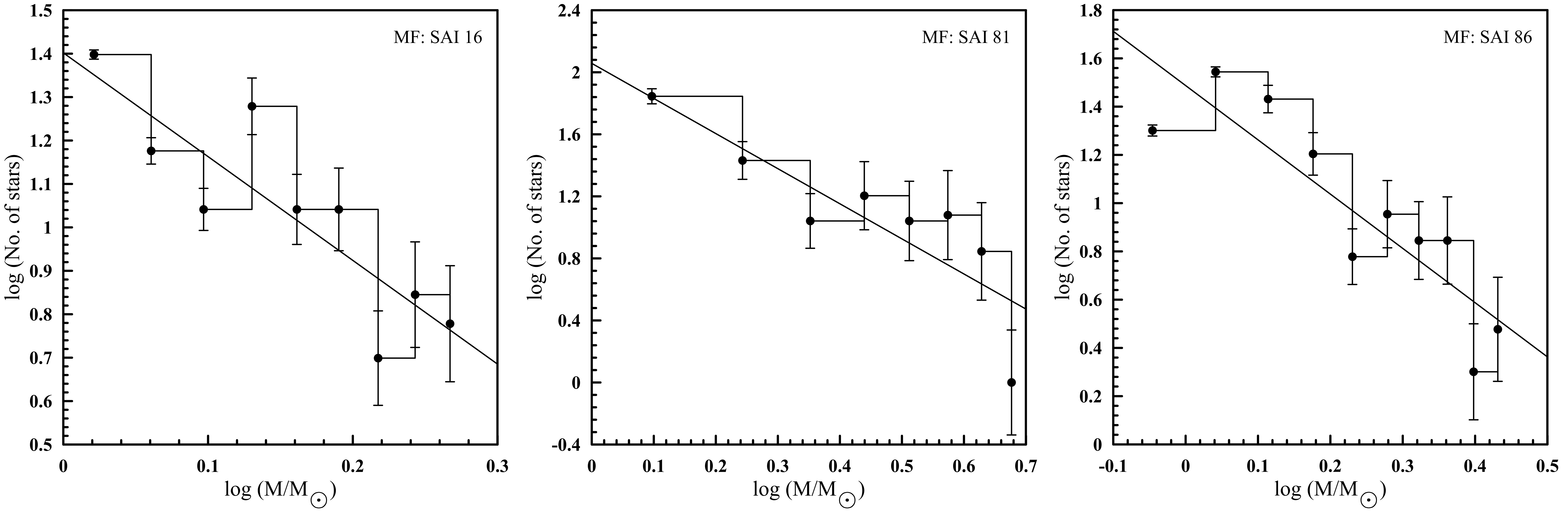}
\caption{Mass function plots for the clusters, fitted with the Salpeter power-law line to illustrate the mass distribution characteristics.}
\label{Fig: 7}
\end{figure*}

The mean mass of the stars in each cluster offers a further understanding of the distribution of stellar masses within the population. To visualize the mass distribution across the clusters, we have plotted the MFs of the three OCs on a logarithmic scale, as shown in Figure \ref{Fig: 7}. The MF slopes ($\alpha$) of the linear fits for each cluster can be derived by using Equation \ref{Eq: 4} and are listed in Table \ref{Tab: 5}, which reflects good agreements with Salpeter's value.

\begin{figure*}
\centering
\includegraphics[width=0.99\linewidth]{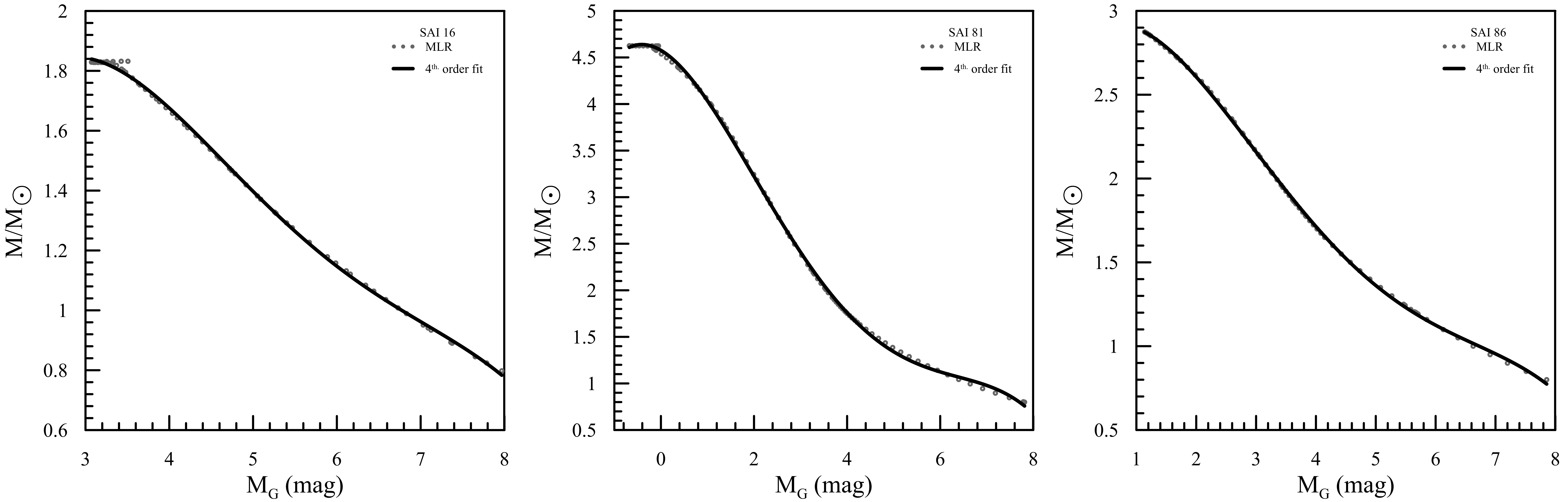}
\caption{Relation between absolute magnitude $M_G$ and mass $(M/M_{\odot})$ from isochrones \citep{2018A&A...616A...4E} of metallicity $Z=0.010\pm0.003,~0.017\pm0.002$, and $0.020\pm0.001$ for SAI 16, SAI 81, and SAI 86, respectively.}
\label{Fig: MLR}
\end{figure*}

\begin{table}[t]%
\centering
\footnotesize
\caption{The MFs and its derived parameters. \label{Tab: 5}}%
\tabcolsep=0pt%
\renewcommand{\arraystretch}{1.3}
\begin{tabular*}{20pc}{@{\extracolsep\fill}lccc@{\extracolsep\fill}}
\hline
\textbf{Parameters} & \textbf{SAI 16}  & \textbf{SAI 81}  & \textbf{SAI 86}  \\
\hline
\hline
$N_{\rm MS}$ & 107  & 155  & 130   \\
MS $M_{\text{C}}$ ($M_{\odot}$) & 142 $\pm$ 12 & 302 $\pm$ 17  & 192 $\pm$ 14   \\
MS $\langle M_C \rangle$ ($M_{\odot}$) & 1.29   & 1.95  & 1.45   \\
$\alpha$ & -2.39$\pm$0.12  & -2.27$\pm$0.10  & -2.50$\pm$0.11  \\
$a_0$ & -1.240$\pm$0.002  & 4.581$\pm$0.005  & 2.800$\pm$0.003  \\
$a_1$ & 2.763$\pm$0.002  & -0.279$\pm$0.002  & 0.383$\pm$0.002  \\
$a_2$ & -0.840$\pm$0.003  & -0.314$\pm$0.003  & -0.340$\pm$0.001  \\
$a_3$ & 0.010$\pm$0.002  & 0.069$\pm$0.002  & 0.056$\pm$0.004  \\
$a_4$ & -0.004$\pm$0.001  & -0.004$\pm$0.001  & -0.003$\pm$0.001  \\
\hline
\end{tabular*}
\end{table}

\subsection{Dynamical and Kinematical Structure}

\subsubsection{The Dynamical Relaxation Time\newline}

The dynamical relaxation time $(T_\text{relax})$ represents the timescale over which a star cluster reaches a state of equilibrium, balancing the opposing forces that drive contraction and possible disruption of the cluster. This time is crucial for understanding the evolutionary state of the cluster, indicating whether it has already undergone significant relaxation or is still in the process of forming and settling \citep{Haroon2017}. A shorter relaxation time suggests that the cluster has had sufficient time to evolve dynamically, while a longer relaxation time may point to a younger, less evolved system.

The dynamical relaxation time is expressed by the following equation \citep{Spitzer1971}:
\begin{equation} \label{Eq: 6}
T_\text{relax} = \frac{8.9 \times 10^5 N^{\frac{1}{2}} R_\text{h}^{\frac{3}{2}}}{\langle M_C \rangle^{\frac{1}{2}} \log(0.4N)},
\end{equation}
where $N$ represents the number of stars within the cluster, $\langle M_C \rangle$ is the mean mass of the stars in the cluster, and $R_{\rm h}$ is the radius (in pc) containing $\sim50\%$ of the cluster mass and can be estimated based on the transformation given in \cite{
2006BaltA..15..547S}. This formula quantifies how the physical properties of the cluster, such as its mass distribution and size, influence the time it takes for the cluster to relax dynamically.
i.e.,
\begin{equation}
R_{\rm h}~=~0.547 \times r_{\rm c} \times \Big(\frac{r_t}{r_{\rm c}}\Big)\textsuperscript{0.486}.
\end{equation}
where $r_{\rm c}$ and $r_{\rm t}$ represent the core and tidal radii, respectively. Based on these parameters, the half-mass radii computed ($R_{\rm h}$) for SAI 16, SAI 81, and SAI 86 are 2.73$\pm$0.61, 4.43$\pm$0.48, and 2.82$\pm$0.60 pc, respectively.

For the clusters under study, the values of $T_\text{relax}$ were calculated as 23 Myr for SAI 16, 42 Myr for SAI 81, and 24 Myr for SAI 86.  These results reflect the varying stages of dynamical evolution across the clusters, with SAI 16 and SAI 86 showing the shortest relaxation time, possibly indicating a more relaxed, older system compared to SAI 81, which appears to be in a more youthful state with a longer relaxation time.

\subsubsection{Convergent Point and Space Velocity\newline}

Stars within a star cluster exhibit a shared motion through space, referred to as proper motion. Due to the perspective effect, these individual motions appear to converge at a specific point on the celestial sphere, commonly known as the apex or the Convergent Point (CP) \citep{Elsanhoury2018}. The process of determining the location of the cluster's apex relies on a technique called the AD diagram, which has been discussed in the works of \cite{Chupina2001} and \cite{Chupina2006}. The AD diagram is a graphical representation that plots the apex positions of individual stars in a cluster, with each star's position being determined by its space velocity components in the ($A_o,~D_o$) coordinate system. The $A_o$ and $D_o$ values correspond to the angular components of the space velocity vector, where $A_o$ denotes the right ascension angle and $D_o$ denotes the declination angle of the apex.

\begin{figure*}[t]
\centering
\includegraphics[width=0.99\linewidth]{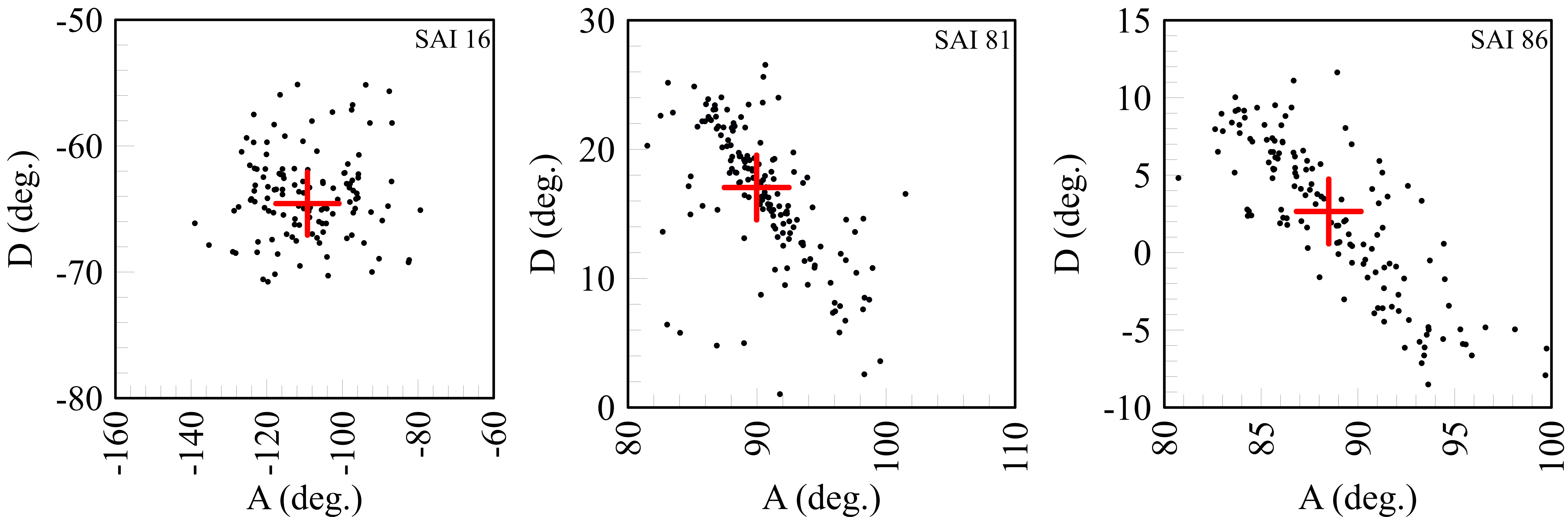}
\caption{The AD-diagram plots for the clusters, with the cross mark indicating the location of the apex point $(A_o,~D_o)$.}
\label{Fig: 8}
\end{figure*}

In this method, the apex is located at the intersection of the proper motion vectors of all the cluster members. The position of the apex ($A_o,~D_o$) in equatorial coordinates can be calculated using the following relations for the right ascension angle $A_o$ and declination angle $D_o$:
\begin{equation} \label{Eq: 7}
A_o = \tan^{-1}\left(\dfrac{\overline{V}_y}{\overline{V}_x}\right),
\end{equation}
\begin{equation} \label{Eq: 8}
D_o = \tan^{-1} \left(\dfrac{\overline{V}_z}{\sqrt{\overline{V}^{2}_{x} + \overline{V}^{2}_{y}}}\right),
\end{equation}
where $V_{x}$, $V_{y}$, and $V_{z}$ are the space velocity components along the respective axes.

For a star in the cluster with known equatorial coordinates $(\alpha,~\delta)$, where $\alpha$ represents the right ascension and $\delta$ the declination, at a distance ($d$) with proper motion components $(\mu_\alpha \cos \delta,~\mu_\delta)$ and radial velocity ($V_r$), as provided by \citet{Hunt2024}, are: -56.26 $\pm$ 13.42 km s$^{-1}$ for SAI 16, 54.97 $\pm$ 0.95 km s$^{-1}$ for SAI 81, and 60.98 $\pm$ 2.47 km s$^{-1}$ for SAI 86. The space velocity components $V_x$, $V_y$, and $V_z$ are computed as follows \citep{Melchior1958}:
\begin{equation} \label{Eq. 9-10-11}
\begin{pmatrix}
V_x \\\\
V_y \\\\
V_z
\end{pmatrix} 
= 
\begin{pmatrix}
-4.74~d~\mu_\alpha\cos{\delta}\sin{\alpha} - 4.74~d~\mu_\delta\sin{\delta}\cos{\alpha} \\
+ V_{\rm r} \cos\delta \cos\alpha \\
+4.74~d~\mu_\alpha\cos{\delta}\sin{\alpha} - 4.74~d~\mu_\delta\sin{\delta}\cos{\alpha} \\
+ V_{\rm r} \cos\delta \cos\alpha \\
+4.74~d~\mu_\delta\cos\delta + V_{\rm r}\sin\delta
\end{pmatrix}
\end{equation}

After calculating the mean values for the space velocity components ($V_x,~V_y,~V_z$), the equatorial coordinates for the CP can be determined. The resulting apex positions for the three studied OCs are illustrated as cross marks in Figure \ref{Fig: 8}. These coordinates, along with additional details, are summarized in Table \ref{Tab: 6}.

To transform the space velocity components from the equatorial frame to Galactic coordinates, we utilized the transformation equations detailed in Eqs. (\ref{Eq. 12}). These transformations allow for the conversion of the velocity components ($V_x,~V_y,~V_z$) into the Galactic coordinate system, represented by $U$, $V$, and $W$, which are the components of the velocity vector relative to the Galactic center. The formulas for calculating these components are given by:
\begin{equation} \label{Eq. 12}
\begin{pmatrix}
U \\\\
V \\\\
W
\end{pmatrix} 
= 
\begin{pmatrix}
-0.0518807421 \; V_{x} -
0.872222642 \; V_{y} -\\
0.4863497200 \; V_{z} \\
+0.4846922369 \; V_{x} -
0.4477920852 \; V_{y} +\\
0.7513692061 \; V_{z}\\
-0.873144899 \; V_{x} -
0.196748341 \; V_{y} +\\
0.4459913295 \; V_{z} \\
\end{pmatrix}
\end{equation}
In addition to the individual velocity components, the mean Galactic space velocities, denoted as $\overline{U}$, $\overline{V}$, and $\overline{W}$, are computed as the mean of the velocity components across all stars in the cluster. These mean velocities are given by the following expressions:
\begin{equation} \label{eq: 15}
\overline{U} = \dfrac{1}{N}\sum ^{N}_{i=1}U_i,\;\;
\overline{V} = \dfrac{1}{N}\sum ^{N}_{i=1}V_i, \;\; \text{and} \; \; \overline{W} = \dfrac{1}{N}\sum ^{N}_{i=1}W_i,
\end{equation}
where $N$ represents the total number of stars in the cluster.

\begin{figure*}[t]
\centering
\includegraphics[width=0.99\linewidth]{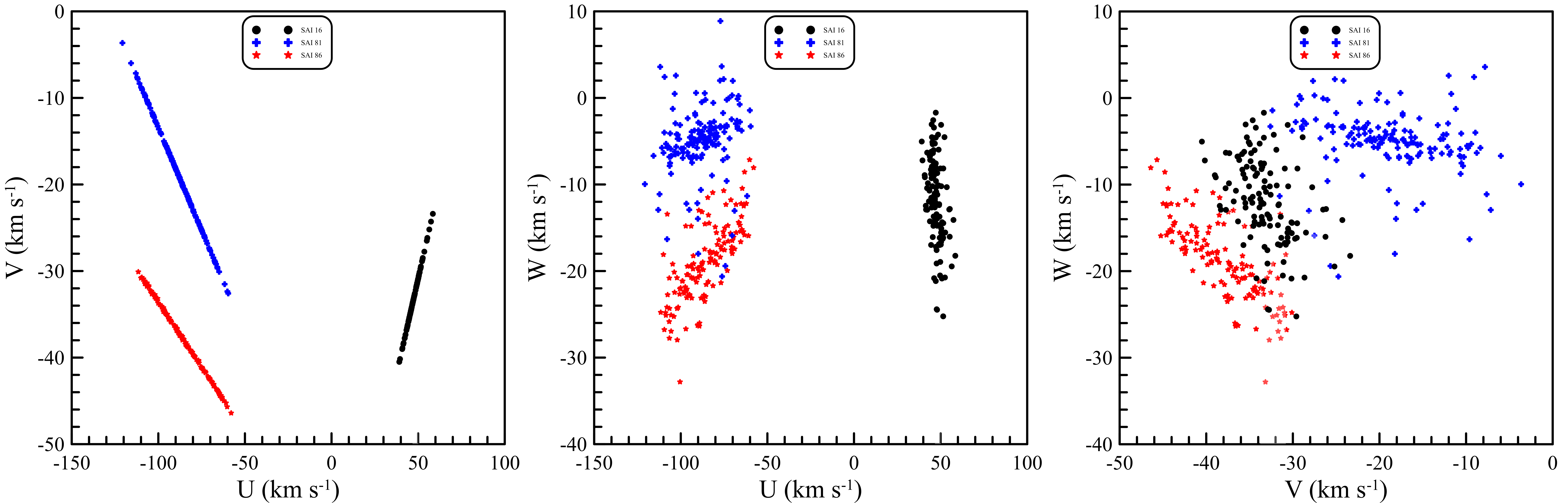}
\caption{The dispersion of the spatial velocity components projected onto the Galactic coordinate system for the clusters under study.}
\label{Fig: 9}
\end{figure*}

The distribution of the spatial velocity components for each Galactic member star within the clusters is shown in Figure \ref{Fig: 9}. Additionally, the mean space velocity components $(U, V, W)$ for the star clusters are summarized in Table \ref{Tab: 6}. 

\subsubsection{Elements of Solar Motion\newline} 

The determination of the solar space velocity components ($U_{\odot},~V_{\odot},~W_{\odot}$) relies on the velocities of a carefully selected set of nearby stars, which serve as a reference for the Sun’s motion within the Galaxy. By utilizing the mean spatial velocity components ($\overline{U},~\overline{V},~\overline{W}$) of a given stellar cluster in Galactic coordinates, the solar space velocity components can be computed in km s$^{-1}$ through the following relations \citep{Elsanhoury2016, Elsanhoury2022, Haroon2025}:
\begin{equation} \label{Eq.16}
U_{\odot} = -\overline{U}, \; V_{\odot} = -\overline{V}, \; \text{and} \; W_{\odot} = -\overline{W}.
\end{equation}
Subsequently, the magnitude of the solar space velocity relative to the observed objects is calculated as:
\begin{equation} \label{Eq. 17}
S_{\odot}=\sqrt{(\overline{U})^2+(\overline{V})^2+(\overline{W})^2}.
\end{equation}
This calculation allows us to determine the Galactic coordinates of the solar apex, defined by the following expressions:
\begin{equation} \label{Eq. 18}
l_\text{A} = \tan^{-1}\left(\frac{-\overline{V}}{\overline{U}}\right) \;\; \text{and} \;\;
b_\text{A} = \sin^{-1} \left(\frac{-\overline{W}}{S_\odot}\right),
\end{equation}
where $l_\text{A}$ and $b_\text{A}$ represent the Galactic longitude and latitude of the solar apex, respectively. Using these relations, we computed the solar space velocity and the solar apex coordinates for each cluster under study, with the results presented in Table \ref{Tab: 6}.

\subsubsection{Orbit Parameters and Galactic Population \newline}
\label{sec:dynnamic}

To investigate the orbital dynamics of star clusters, we employed the {\sc MWPotential2014} framework implemented within the robust {\sc galpy}\footnote{https://galpy.readthedocs.io/en/v1.5.0/} package, a Python-based toolkit extensively utilized in Galactic dynamics research. This model, as introduced and elaborated by \citet{Bovy2015}, offers a comprehensive representation of the gravitational potential of the MW. The parameters integrated into the model enable precise simulations, providing a foundation for examining the kinematic and structural properties of stellar populations.

\begin{figure*}
\centering
\includegraphics[width=0.9\linewidth]{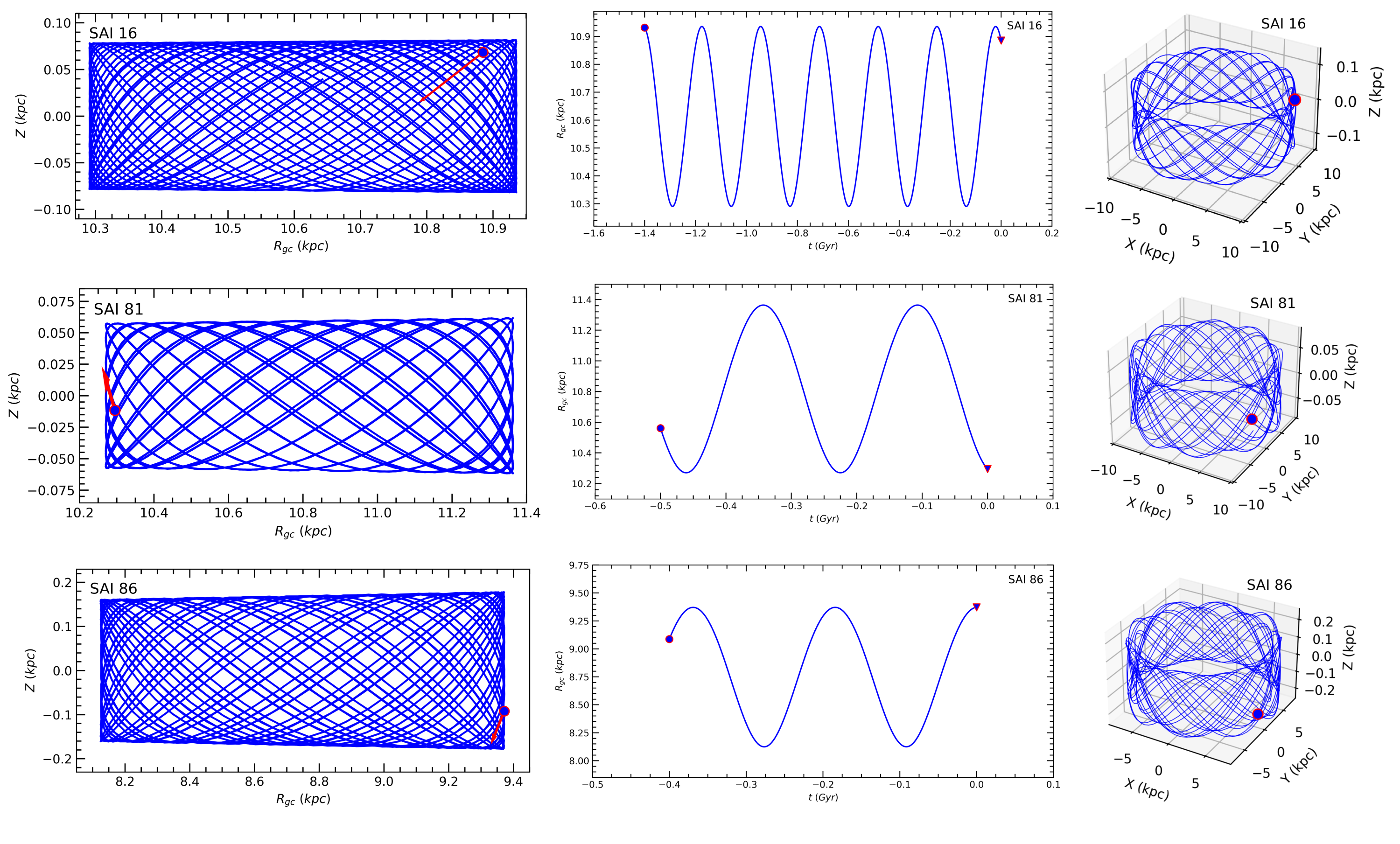}
\caption{The Galactic orbits and birth radii of the SAI clusters are depicted on three separate planes: $Z$ × $R_{\rm gc}$ (left), $R_{\rm gc}$ × $t$ (middle), and $X$ × $Y$ × $Z$ (right). Red circles mark the present locations of the SAI OCs, while triangles indicate their original formation positions. Red arrows represent the motion vectors of the clusters.}
\label{Fig: 10}
\end{figure*}

The methodology adopted here builds on established approaches previously validated through analyses of both individual stellar objects and OCs \citep{Tasdemir2023, Yucel2024,Tasdemir2025a, Tasdemir2025b}. Key Galactic parameters, such as the Galactocentric distance of the Sun ($R_{\rm o} = 8.20 \pm 0.10$ kpc) and its rotational velocity ($V_{\rm rot} = 220$ km s$^{-1}$), were central to our calculations, as detailed in the Method and Results Section. Additionally, the Sun's vertical offset from the Galactic plane was set at $Z_{\rm 0} = 25 \pm 5$ pc, consistent with the findings of \citet{Juric2008}. These parameters ensure a realistic representation of the MW's structural and dynamic characteristics.

To investigate the dynamical orbits of the clusters SAI 16, SAI 81, and SAI 86, it is essential to consider key parameters such as their equatorial coordinates ($\alpha,~\delta$), distances ($d$), and proper motion components ($\mu_\alpha\cos\delta,~\mu_\delta$). Radial velocity ($V_{\rm r}$) data \citep{Hunt2024} also play a crucial role in these calculations. These parameters were utilized to determine the current Galactic positions of the clusters. Then, numerical orbit integration was performed to trace their trajectories, with a time step of 1 Myr, corresponding to their estimated ages \citep{Dursun2024, Elsanhoury2024, Haroon_2025}. The results, summarized in Table \ref{Tab: 6} and visualized in Figure \ref{Fig: 10}, reveal that the clusters exhibit nearly circular orbital paths. Their vertical distances from the Galactic plane suggest a strong association with the young stellar disk of the MW. These findings align with the classification criteria established by \citet{Ak2015}.

\begin{table}
%\tiny
\centering
\renewcommand{\arraystretch}{1.4}
\setlength{\tabcolsep}{2.8pt}
\caption{The evolving, kinematical, and dynamical parameters for the SAI 16 SAI 81, and SAI 86 clusters.}
%\footnotesize
\begin{tabular}{l|cccc}
\hline
\textbf{Parameter} & \textbf{SAI 16} & \textbf{SAI 81} & \textbf{SAI 86} \\
\hline
\hline
& \multicolumn{3}{c}{Evolving Parameters} \\
\hline
\hline
$R_h$ (pc) & 2.73$\pm$0.61 & 4.43$\pm$0.48&2.82$\pm$0.60\\
$T_{relax}$ (Myr) & 23 & 42 & 24\\
$\tau$ & 59 & 3 & 19\\
$\tau_{ev}$ (Myr) & 2300 & 4200 & 2400\\
\hline
& \multicolumn{3}{c}{Kinematical Parameters} \\
\hline
\hline
$\overline{V_x}$ (km s$^{-1}$) & -8.19$\pm$2.86 & -0.32$\pm$0.01 & 2.68$\pm$0.06\\
$\overline{V_y}$ (km s$^{-1}$) & -24.17$\pm$4.92 & 86.47$\pm$9.30 & 95.51$\pm$9.77\\
$\overline{V_z}$ (km s$^{-1}$) & -53.27$\pm$7.30 & 26.46$\pm$4.14 &4.81$\pm$0.46\\
$\overline{U}$ (km s$^{-1}$)& ~47.42$\pm$6.89 & -88.27$\pm$9.40 & -85.79$\pm$9.26\\
$\overline{V}$ (km s$^{-1}$)& -33.17$\pm$5.76 & -18.99$\pm$4.36 & -37.86$\pm$6.15\\
$\overline{W}$ (km s$^{-1}$)& -11.85$\pm$3.44 & -4.94$\pm$0.45 & -18.98$\pm$4.35\\
$S_{\odot}$ (km s$^{-1}$)&  ~59.07$\pm$7.69& ~90.43$\pm$9.51 &~95.67$\pm$9.78\\
\hline
& \multicolumn{3}{c}{Dynamical Parameters} \\
\hline
\hline
$A_o~(^o)$ & -108.72$\pm$0.10 & 90.21$\pm$0.10 & 88.40$\pm$0.11 \\
$D_o~(^o)$ & -64.41$\pm$0.13 & 17.02$\pm$0.37 &2.88$\pm$0.01\\ 
$l_A~(^o)$ &34.98&-12.14&-23.81\\
$b_A~(^o)$ &~11.57&~3.13&11.45\\
$Z_{\rm max}$ (kpc) &0.08$\pm$0.01 &  0.06$\pm$0.08  &   0.18$\pm$0.03 \\
$R_{\rm a}$ (kpc) &  10.93$\pm$0.22 &  11.36$\pm$0.07 &  9.37$\pm$0.02  \\
$R_{\rm p}$ (kpc) &  10.29$\pm$0.78 &  10.27$\pm$0.08    &  8.12$\pm$0.11  \\
$R_{\rm m}$ (kpc) &  10.61$\pm$0.50 &  10.81$\pm$0.07    &  8.75$\pm$0.07  \\
$e$              &  0.03$\pm$0.02 &  0.05$\pm$0.01  &  0.07$\pm$0.01  \\
$T_{\rm p}$ (Myr) & 306$\pm$16 &  313$\pm$1  &  247$\pm$2  \\
\hline
\end{tabular}
\label{Tab: 6}
\end{table}

\section{Discussion}\label{sec4}

SAI 16, SAI 81, and SAI 86 OCs have remained largely unexplored in the literature, with no dedicated studies providing a comprehensive analysis of their fundamental properties. In this study, we conducted a comprehensive analysis of these clusters, examining their fundamental astrophysical properties, kinematics, dynamics, and Galactic context.

As summarized in Table \ref{Tab: 1}, previous studies report age ($\log t$) estimates ranging from 8.46 to 9 for SAI 16, 8.13 to 8.37 for SAI 81, and 8.54 to 9.01 for SAI 86. Our analysis yields age values of 9.13 for SAI 16, 8.10 for SAI 81, and 8.65 for SAI 86. Although the results for SAI 16 and SAI 81 are within the boundaries of previously reported ranges, the estimated age for SAI 86 is in agreement with the literature. The limited number of studies available on these clusters, most of which are based on $Gaia$ DR2 data, and the inherent degeneracies in the applied models have contributed to the discrepancies and broad uncertainties in the reported age values. Using $Gaia$ DR3 data and the ASteCA code, we identified the most probable members of the cluster and derived refined astrometric and astrophysical parameters. These results improve the precision of cluster parameters, refining their astrophysical and kinematical properties within the galactic framework and contributing to the existing literature.

Table \ref{Tab: 4} highlights the continuous advancements in the refinement of the fundamental parameters of these OCs. Although the results of the current study generally corroborate the previous findings, some variations are evident, probably due to differences in study methodologies, datasets, or analytical techniques. These disparities between our results and the literature values further underscore the importance of ongoing refinement and the necessity for improved precision when characterizing these stellar systems. This study presents, for the first time, the MFs of the OCs SAI 16, SAI 81, and SAI 86, filling a significant gap in the literature. The newly derived MFs enable a detailed examination of the clusters' stellar distributions and evolutionary stages, providing a basis for future comparative studies.

Open clusters undergo dynamical evolution, leading to central concentration as massive stars migrate inward while low-mass stars move outward and eventually escape. The $R_{\rm h}$/$r_{\rm t}$ ratio, defined as the half-mass radius to tidal radius, is a key parameter for assessing tidal interactions and stability of the cluster. Lower $R_{\rm h}$/$r_{\rm t}$ values indicate more compact clusters, which are less prone to tidal stripping, particularly in regions of strong Galactic gravitational influence \citep{Karatas2023}.

\citet{Maurya2023} estimated $R_{\rm h}$/$r_{\rm t}$ ratios using total cluster masses derived from the \citet{Kroupa2001} MF, incorporating faint stars down to 0.08 $M_{\odot}$. Their results reveal a positive correlation between $R_{\rm h}$/$r_{\rm t}$ and $R_{\rm gc}$, with a slope of 0.06 $\pm$ 0.01 and $r^2$ = 0.93, indicating that clusters at larger $R_{\rm gc}$ tend to have higher $R_{\rm h}$/$r_{\rm t}$ values. This trend is consistent with this study, whose results suggest that weaker external gravitational forces at larger $R_{\rm gc}$ allow clusters to fill a greater fraction of their tidal volume. We have calculated the $R_{\rm h}$/$r_{\rm t}$ values for the SAI 16, SAI 81, and SAI 86 clusters, which are 0.419, 0.455, and 0.377, respectively. Our findings align with this interpretation, although the small and age-diverse sample in this study underscores the need for further investigations with larger, age-homogeneous cluster samples across various Galactic environments.

The dynamical state of clusters can be characterized by computing the dynamical evolution parameter, defined as $ \tau = \text{age} / T_{\text{relax}} $. For all clusters analyzed, this parameter is significantly greater than unity ($ \tau \gg 1 $), indicating that they have reached a dynamically relaxed state. The evaporation time, given by $ \tau_{\text{ev}} \simeq 10^2 T_{\text{relax}}$ (in Myr), represents the timescale over which all member stars are ejected due to internal stellar encounters \citep{2001ApJ...553..744A}. In this process, low-mass stars predominantly escape at low velocities through the Lagrange points \citep{2008MNRAS.387.1248K}.

In this study, the mean Galactocentric radius, $R_{\rm m}$ = ($R_{\rm a}$ + $R_{\rm p}$)/2, also referred to as the guiding or mean orbital radius, was calculated for each of the clusters. The rotational velocities, eccentricities, and orbital angular momentum values suggest that SAI 16, SAI 81, and SAI 86 exhibit characteristics typical of the Galactic thin disk. These clusters follow nearly circular orbits around the Galactic center, with eccentricities ranging from 0.03 to 0.07, indicating their location near the Galactic disc, where they are likely influenced by tidal forces. Based on their revolution periods, SAI 16, SAI 81, and SAI 86 have completed 306, 313, and 247 orbits around the Galactic center, respectively (see Table \ref{Tab: 6}).

The orbital parameters reported by \citet{Tarricq2022} for these clusters include apo- peri center, eccentricity, and maximum vertical distance from the disc, with values for SAI 16 of $R_{\rm a}$ = 11.213 kpc, $R_{\rm p}$ = 13.774 kpc, $e$ = 0.103, and $Z_{\rm max}$ = 0.135 kpc; for SAI 81, $R_{\rm a}$ = 10.765 kpc, $R_{\rm p}$ = 11.709 kpc, $e$ = 0.042, and $Z_{\rm max}$ = 0.052 kpc; and for SAI 86, $R_{\rm a}$ = 8.922 kpc, $R_{\rm p}$ = 9.899 kpc, $e$ = 0.052, and $Z_{\rm max}$ = 0.219 kpc. These values differ from those obtained in this study due to variations in the adopted radial velocities for the clusters. The current work uses updated values derived from $Gaia$ DR3 data, as provided by \citet{Hunt2024}, which offer more accurate and reliable parameters for further analysis.

\section{Summary}\label{sec5}
\begin{itemize}
\item In this study, we conducted comprehensive photometric, astrometric, kinematic, and orbital investigations of the OCs SAI 16, SAI 81, and SAI 86 using \textit{Gaia} DR3 data, supplemented by the ASteCA code. The ASteCA tool facilitated the determination of various cluster parameters, including centers, RDPs, membership probabilities, and astrophysical parameters derived from CMDs.

\item We analyzed the coordinates of the centers of the three clusters, with the results presented in both equatorial and Galactic coordinate systems, as shown in Table \ref{Tab: 2}. Subsequently, the RDP of each cluster was fitted using a King profile model to estimate their core, limiting, and tidal radii. The core radii derived ranged from 3.41 arcmin for SAI 86 to 5.94 arcmin for SAI 81, and the limiting radii extended from 3.63 arcmin to 4.78 arcmin, as detailed in Table \ref{Tab: 3}.

\item The construction of CMDs using \textit{Gaia} photometric magnitudes allowed for the fitting of theoretical isochrones from the PARSEC v1.2S model, from which we derived important astrophysical parameters such as age, metallicity, distance, and reddening. A comparison of these parameters with those found in the literature revealed both strong consistencies and a few discrepancies, as illustrated in Table \ref{Tab: 4}.

\item Moreover, the distance of each cluster from the Galactic center and their positions within the Milky Way (MW) were determined based on the derived distances and Galactic coordinates. Remarkably, all three clusters showed a negative projected distance towards the Galactic plane ($X_\odot$), suggesting that they are located on the side of the Milky Way opposite the Galactic center, as viewed from the Sun.

\item The study also involved the construction of luminosity and mass functions (LFs and MFs) for the clusters. The LF was derived from the absolute magnitudes of the cluster members, and the MF was obtained by converting the luminosities to masses using a mass-luminosity relation (MLR). The mass function slopes for SAI 16, SAI 81, and SAI 86 were found to be -2.39 $\pm$ 0.12, -2.27 $\pm$ 0.10, and -2.50 $\pm$ 0.11, respectively, which are in good agreement with the slope proposed by \citet{Salpeter1955}.

\item Furthermore, we examined the dynamical status of the OCs by assessing the relaxation times. Our analysis of the dynamical relaxation state indicated that all three clusters have reached a dynamical equilibrium state. The convergent point (CP) of each cluster was also determined using the AD diagram technique, with the apexes of the cluster members plotted according to their velocity components.

\item In addition, the space velocity components $(U,~V,~W)$ of the clusters were estimated in the Galactic coordinate system using equatorial-to-Galactic transformation formulas. The solar elements for each cluster were also derived. Finally, the open clusters SAI 16, SAI 81, and SAI 86 were found to exhibit maximum distances from the Galactic plane of 0.08 $\pm$ 0.01 kpc, 0.06 $\pm$ 0.08 kpc, and 0.18 $\pm$ 0.03 kpc, respectively. These distances strongly support their classification as members of the young stellar disk population, further enriching our understanding of their positions and dynamics within the MW.
\end{itemize}

\section*{ACKNOWLEDGEMENTS}
We express our gratitude to the anonymous referee for their valuable comments and suggestions, which are very helpful in improving our manuscript. This work presents results from the European Space Agency (ESA) space mission Gaia. $Gaia$ data are being processed by the $Gaia$ Data Processing and Analysis Consortium (DPAC). Funding for the DPAC is provided by national institutions, in particular, the institutions participating in the $Gaia$ Multi-Lateral Agreement (MLA). The $Gaia$ mission website is \url{https://www.cosmos.esa.int/gaia}. The $Gaia$ archive website is \url{https://archives.esac.esa.int/gaia}. The authors would like to express their gratitude to the Deanship of Scientific Research at Northern Border University, Arar, KSA, for funding this research under project number "NBU-FFR-2025-237-05".

\bibliography{JAA}

\begin{thebibliography}{}
\expandafter\ifx\csname natexlab\endcsname\relax\def\natexlab#1{#1}\fi

\bibitem[{{Adams} \& {Myers}(2001)}]{2001ApJ...553..744A}
{Adams}, F.~C., \& {Myers}, P.~C. 2001, \apj, 553, 744

\bibitem[{{Ak} {$et~al$.}(2015){Ak}, {Bilir}, {{\"O}zd{\"o}nmez}, {Soydugan}, {Soydugan}, {P{\"u}sk{\"u}ll{\"u}}, {Ak}, \& {Eker}}]{Ak2015}
{Ak}, T., {Bilir}, S., {{\"O}zd{\"o}nmez}, A., {$et~al$.} 2015, Astrophysics and Space Science, 357, 72

\bibitem[{Akbulut {$et~al$.}(2021)Akbulut, Ak, Yontan, Bilir, Ak, Banks, Kaan~Ulgen, \& Paunzen}]{Akbulut2021}
Akbulut, B., Ak, S., Yontan, T., {$et~al$.} 2021, Astrophysics and Space Science, 366, 68

\bibitem[{Badawy {$et~al$.}(2023)Badawy, Tadross, Hendy, Ismail, \& Mouner}]{Badawy2023}
Badawy, W., Tadross, A.~L., Hendy, Y.~H., Ismail, M.~N., \& Mouner, A. 2023, Revista Mexicana de Astronom{\i}a y Astrof{\i}sica, 59, 155

\bibitem[{Bisht {$et~al$.}(2019)Bisht, Yadav, Ganesh, Durgapal, Rangwal, \& Fynbo}]{Bisht2019}
Bisht, D., Yadav, R., Ganesh, S., {$et~al$.} 2019, Monthly Notices of the Royal Astronomical Society, 482, 1471

\bibitem[{Bland-Hawthorn {$et~al$.}(2019)Bland-Hawthorn, Sharma, Tepper-Garcia, Binney, Freeman, Hayden, Kos, De~Silva, Ellis, Lewis, {$et~al$.}}]{Bland2019}
Bland-Hawthorn, J., Sharma, S., Tepper-Garcia, T., {$et~al$.} 2019, Monthly Notices of the Royal Astronomical Society, 486, 1167

\bibitem[{{Bonatto} \& {Bica}(2009)}]{BonattoBica2009}
{Bonatto}, C., \& {Bica}, E. 2009, \mnras, 397, 1915

\bibitem[{Bonatto {$et~al$.}(2006)Bonatto, Kerber, Bica, \& Santiago}]{Bonatto2006}
Bonatto, C., Kerber, L., Bica, E., \& Santiago, B.~X. 2006, Astronomy \& Astrophysics, 446, 121

\bibitem[{{Bouma} {$et~al$.}(2019){Bouma}, {Hartman}, {Bhatti}, {Winn}, \& {Bakos}}]{Bouma2019}
{Bouma}, L.~G., {Hartman}, J.~D., {Bhatti}, W., {Winn}, J.~N., \& {Bakos}, G.~{\'A}. 2019, \apjs, 245, 13

\bibitem[{{Bovy}(2015)}]{Bovy2015}
{Bovy}, J. 2015, The Astrophysical Journal, 216, 29

\bibitem[{Bressan {$et~al$.}(2012)Bressan, Marigo, Girardi, Salasnich, Dal~Cero, Rubele, \& Nanni}]{Bressan2012}
Bressan, A., Marigo, P., Girardi, L., {$et~al$.} 2012, Monthly Notices of the Royal Astronomical Society, 427, 127

\bibitem[{Cantat-Gaudin {$et~al$.}(2018)Cantat-Gaudin, Jordi, Vallenari, Bragaglia, Balaguer-N{\'u}{\~n}ez, Soubiran, Bossini, Moitinho, Castro-Ginard, Krone-Martins, {$et~al$.}}]{Cantat-Gaudin2018}
Cantat-Gaudin, T., Jordi, C., Vallenari, A., {$et~al$.} 2018, Astronomy \& Astrophysics, 618, A93

\bibitem[{Cantat-Gaudin {$et~al$.}(2020)Cantat-Gaudin, Anders, Castro-Ginard, Jordi, Romero-G{\'o}mez, Soubiran, Casamiquela, Tarricq, Moitinho, Vallenari, {$et~al$.}}]{CantatGaudin2020}
Cantat-Gaudin, T., Anders, F., Castro-Ginard, A., {$et~al$.} 2020, Astronomy \& Astrophysics, 640, A1

\bibitem[{Carraro {$et~al$.}(2017)Carraro, Turner, Majaess, Baume, Gamen, \& Lera}]{Carraro2017}
Carraro, G., Turner, D.~G., Majaess, D.~J., {$et~al$.} 2017, The Astronomical Journal, 153, 156

\bibitem[{Carraro {$et~al$.}(2008)Carraro, V{\'a}zquez, \& Moitinho}]{Carraro2008}
Carraro, G., V{\'a}zquez, R.~A., \& Moitinho, A. 2008, Astronomy \& Astrophysics, 482, 777

\bibitem[{{Carrera} {$et~al$.}(2019){Carrera}, {Bragaglia}, {Cantat-Gaudin}, {Vallenari}, {Balaguer-N{\'u}{\~n}ez}, {Bossini}, {Casamiquela}, {Jordi}, {Sordo}, \& {Soubiran}}]{2019A&A...623A..80C}
{Carrera}, R., {Bragaglia}, A., {Cantat-Gaudin}, T., {$et~al$.} 2019, \aap, 623, A80

\bibitem[{Casagrande \& VandenBerg(2018)}]{Casagrande2018}
Casagrande, L., \& VandenBerg, D.~A. 2018, Monthly Notices of the Royal Astronomical Society: Letters, 479, L102

\bibitem[{Castro-Ginard {$et~al$.}(2021)Castro-Ginard, McMillan, Luri, Jordi, Romero-G{\'o}mez, Cantat-Gaudin, Casamiquela, Tarricq, Soubiran, \& Anders}]{Castro-Ginard2021}
Castro-Ginard, A., McMillan, P.~J., Luri, X., {$et~al$.} 2021, Astronomy \& Astrophysics, 652, A162

\bibitem[{Castro-Ginard {$et~al$.}(2022)Castro-Ginard, Jordi, Luri, Cantat-Gaudin, Carrasco, Casamiquela, Anders, Balaguer-Nu{\~n}ez, \& Badia}]{Castro2022}
Castro-Ginard, A., Jordi, C., Luri, X., {$et~al$.} 2022, Astronomy \& Astrophysics, 661, A118

\bibitem[{{Cavallo} {$et~al$.}(2024){Cavallo}, {Spina}, {Carraro}, {Magrini}, {Poggio}, {Cantat-Gaudin}, {Pasquato}, {Lucatello}, {Ortolani}, \& {Schiappacasse-Ulloa}}]{Cavallo2024}
{Cavallo}, L., {Spina}, L., {Carraro}, G., {$et~al$.} 2024, \aj, 167, 12

\bibitem[{{{\c{C}}{\i}nar} {$et~al$.}(2024){{\c{C}}{\i}nar}, {Ta{\c{s}}demir}, {Koc}, \& {Iyer}}]{Dursun2024}
{{\c{C}}{\i}nar}, D.~C., {Ta{\c{s}}demir}, S., {Koc}, S., \& {Iyer}, S. 2024, Physics and Astronomy Reports, 2, 1

\bibitem[{Chen {$et~al$.}(2003)Chen, Hou, \& Wang}]{Chen2003}
Chen, L., Hou, J., \& Wang, J. 2003, The Astronomical Journal, 125, 1397

\bibitem[{{Chupina} {$et~al$.}(2001){Chupina}, {Reva}, \& {Vereshchagin}}]{Chupina2001}
{Chupina}, N.~V., {Reva}, V.~G., \& {Vereshchagin}, S.~V. 2001, \aap, 371, 115

\bibitem[{{Chupina, N. V.} {$et~al$.}(2006){Chupina, N. V.}, {Reva, V. G.}, \& {Vereshchagin, S. V.}}]{Chupina2006}
{Chupina, N. V.}, {Reva, V. G.}, \& {Vereshchagin, S. V.} 2006, \aap, 451, 909

\bibitem[{Dias \& L{\'e}pine(2005)}]{Dias2005}
Dias, W.~S., \& L{\'e}pine, J. R.~D. 2005, The Astrophysical Journal, 629, 825

\bibitem[{Dias {$et~al$.}(2021)Dias, Monteiro, Moitinho, L{\'e}pine, Carraro, Paunzen, Alessi, \& Villela}]{Dias2021}
Dias, W.~S., Monteiro, H., Moitinho, A., {$et~al$.} 2021, Monthly Notices of the Royal Astronomical Society, 504, 356

\bibitem[{Elsanhoury \& Amin(2019)}]{Elsanhoury2019}
Elsanhoury, W., \& Amin, M.~Y. 2019, Serbian Astronomical Journal, 45

\bibitem[{Elsanhoury {$et~al$.}(2022)Elsanhoury, Amin, Haroon, \& Awad}]{Elsanhoury2022}
Elsanhoury, W., Amin, M.~Y., Haroon, A., \& Awad, Z. 2022, Journal of Astrophysics and Astronomy, 43, 26

\bibitem[{Elsanhoury {$et~al$.}(2016)Elsanhoury, Haroon, Chupina, Vereshchagin, Sariya, Yadav, \& Jiang}]{Elsanhoury2016}
Elsanhoury, W., Haroon, A., Chupina, N., {$et~al$.} 2016, New Astronomy, 49, 32

\bibitem[{Elsanhoury {$et~al$.}(2018)Elsanhoury, Postnikova, Chupina, Vereshchagin, Sariya, Yadav, \& Jiang}]{Elsanhoury2018}
Elsanhoury, W., Postnikova, E., Chupina, N., {$et~al$.} 2018, Astrophysics and space science, 363, 1

\bibitem[{{Elsanhoury} {$et~al$.}(2025){Elsanhoury}, {Haroon}, {Elkholy}, \& {{\c{C}}{\i}nar}}]{Elsanhoury2024}
{Elsanhoury}, W.~H., {Haroon}, A.~A., {Elkholy}, E.~A., \& {{\c{C}}{\i}nar}, D.~C. 2025, Journal of Astrophysics and Astronomy, 46, 21

\bibitem[{{Evans} {$et~al$.}(2018){Evans}, {Riello}, {De Angeli}, {Carrasco}, {Montegriffo}, {Fabricius}, {Jordi}, {Palaversa}, {Diener}, {Busso}, {Cacciari}, {van Leeuwen}, {Burgess}, {Davidson}, {Harrison}, {Hodgkin}, {Pancino}, {Richards}, {Altavilla}, {Balaguer-N{\'u}{\~n}ez}, {Barstow}, {Bellazzini}, {Brown}, {Castellani}, {Cocozza}, {De Luise}, {Delgado}, {Ducourant}, {Galleti}, {Gilmore}, {Giuffrida}, {Holl}, {Kewley}, {Koposov}, {Marinoni}, {Marrese}, {Osborne}, {Piersimoni}, {Portell}, {Pulone}, {Ragaini}, {Sanna}, {Terrett}, {Walton}, {Wevers}, \& {Wyrzykowski}}]{2018A&A...616A...4E}
{Evans}, D.~W., {Riello}, M., {De Angeli}, F., {$et~al$.} 2018, \aap, 616, A4

\bibitem[{Ferreira {$et~al$.}(2021)Ferreira, Corradi, Maia, Angelo, \& Santos~Jr}]{Ferreira2021}
Ferreira, F.~A., Corradi, W., Maia, F., Angelo, M., \& Santos~Jr, J. 2021, Monthly Notices of the Royal Astronomical Society: Letters, 502, L90

\bibitem[{Friel(2013)}]{Friel2013}
Friel, E.~D. 2013, Planets, Stars and Stellar Systems. Volume 5: Galactic Structure and Stellar Populations, 5, 347

\bibitem[{{Gaia Collaboration} {$et~al$.}(2021){Gaia Collaboration}, {Brown}, {Vallenari}, {Prusti}, {de Bruijne}, {Babusiaux}, {Biermann}, {Creevey}, {Evans}, {Eyer}, {Hutton}, {Jansen}, {Jordi}, {Klioner}, {Lammers}, {Lindegren}, {Luri}, {Mignard}, {Panem}, {Pourbaix}, {Randich}, {Sartoretti}, {Soubiran}, {Walton}, {Arenou}, {Bailer-Jones}, {Bastian}, {Cropper}, {Drimmel}, {Katz}, {Lattanzi}, {van Leeuwen}, {Bakker}, {Cacciari}, {Casta{\~n}eda}, {De Angeli}, {Ducourant}, {Fabricius}, {Fouesneau}, {Fr{\'e}mat}, {Guerra}, {Guerrier}, {Guiraud}, {Jean-Antoine Piccolo}, {Masana}, {Messineo}, {Mowlavi}, {Nicolas}, {Nienartowicz}, {Pailler}, {Panuzzo}, {Riclet}, {Roux}, {Seabroke}, {Sordo}, {Tanga}, {Th{\'e}venin}, {Gracia-Abril}, {Portell}, {Teyssier}, {Altmann}, {Andrae}, {Bellas-Velidis}, {Benson}, {Berthier}, {Blomme}, {Brugaletta}, {Burgess}, {Busso}, {Carry}, {Cellino}, {Cheek}, {Clementini}, {Damerdji}, {Davidson}, {Delchambre}, {Dell'Oro}, {Fern{\'a}ndez-Hern{\'a}ndez}, {Galluccio}, {Garc{\'\i}a-Lario},
  {Garcia-Reinaldos}, {Gonz{\'a}lez-N{\'u}{\~n}ez}, {Gosset}, {Haigron}, {Halbwachs}, {Hambly}, {Harrison}, {Hatzidimitriou}, {Heiter}, {Hern{\'a}ndez}, {Hestroffer}, {Hodgkin}, {Holl}, {Jan{\ss}en}, {Jevardat de Fombelle}, {Jordan}, {Krone-Martins}, {Lanzafame}, {L{\"o}ffler}, {Lorca}, {Manteiga}, {Marchal}, {Marrese}, {Moitinho}, {Mora}, {Muinonen}, {Osborne}, {Pancino}, {Pauwels}, {Petit}, {Recio-Blanco}, {Richards}, {Riello}, {Rimoldini}, {Robin}, {Roegiers}, {Rybizki}, {Sarro}, {Siopis}, {Smith}, {Sozzetti}, {Ulla}, {Utrilla}, {van Leeuwen}, {van Reeven}, {Abbas}, {Abreu Aramburu}, {Accart}, {Aerts}, {Aguado}, {Ajaj}, {Altavilla}, {{\'A}lvarez}, {{\'A}lvarez Cid-Fuentes}, {Alves}, {Anderson}, {Anglada Varela}, {Antoja}, {Audard}, {Baines}, {Baker}, {Balaguer-N{\'u}{\~n}ez}, {Balbinot}, {Balog}, {Barache}, {Barbato}, {Barros}, {Barstow}, {Bartolom{\'e}}, {Bassilana}, {Bauchet}, {Baudesson-Stella}, {Becciani}, {Bellazzini}, {Bernet}, {Bertone}, {Bianchi}, {Blanco-Cuaresma}, {Boch}, {Bombrun}, {Bossini},
  {Bouquillon}, {Bragaglia}, {Bramante}, {Breedt}, {Bressan}, {Brouillet}, {Bucciarelli}, {Burlacu}, {Busonero}, {Butkevich}, {Buzzi}, {Caffau}, {Cancelliere}, {C{\'a}novas}, {Cantat-Gaudin}, {Carballo}, {Carlucci}, {Carnerero}, {Carrasco}, {Casamiquela}, {Castellani}, {Castro-Ginard}, {Castro Sampol}, {Chaoul}, {Charlot}, {Chemin}, {Chiavassa}, {Cioni}, {Comoretto}, {Cooper}, {Cornez}, {Cowell}, {Crifo}, {Crosta}, {Crowley}, {Dafonte}, {Dapergolas}, {David}, \& {David}}]{GaiaEDR3}
{Gaia Collaboration}, {Brown}, A.~G.~A., {Vallenari}, A., {$et~al$.} 2021, \aap, 649, A1

\bibitem[{{Gaia Collaboration} {$et~al$.}(2023){Gaia Collaboration}, {Vallenari}, {Brown}, {Prusti}, {de Bruijne}, {Arenou}, {Babusiaux}, {Biermann}, {Creevey}, {Ducourant}, {Evans}, {Eyer}, {Guerra}, {Hutton}, {Jordi}, {Klioner}, {Lammers}, {Lindegren}, {Luri}, {Mignard}, {Panem}, {Pourbaix}, {Randich}, {Sartoretti}, {Soubiran}, {Tanga}, {Walton}, {Bailer-Jones}, {Bastian}, {Drimmel}, {Jansen}, {Katz}, {Lattanzi}, {van Leeuwen}, {Bakker}, {Cacciari}, {Casta{\~n}eda}, {De Angeli}, {Fabricius}, {Fouesneau}, {Fr{\'e}mat}, {Galluccio}, {Guerrier}, {Heiter}, {Masana}, {Messineo}, {Mowlavi}, {Nicolas}, {Nienartowicz}, {Pailler}, {Panuzzo}, {Riclet}, {Roux}, {Seabroke}, {Sordo{\o}rcit}, {Th{\'e}venin}, {Gracia-Abril}, {Portell}, {Teyssier}, {Altmann}, {Andrae}, {Audard}, {Bellas-Velidis}, {Benson}, {Berthier}, {Blomme}, {Burgess}, {Busonero}, {Busso}, {C{\'a}novas}, {Carry}, {Cellino}, {Cheek}, {Clementini}, {Damerdji}, {Davidson}, {de Teodoro}, {Nu{\~n}ez Campos}, {Delchambre}, {Dell'Oro}, {Esquej},
  {Fern{\'a}ndez-Hern{\'a}ndez}, {Fraile}, {Garabato}, {Garc{\'\i}a-Lario}, {Gosset}, {Haigron}, {Halbwachs}, {Hambly}, {Harrison}, {Hern{\'a}ndez}, {Hestroffer}, {Hodgkin}, {Holl}, {Jan{\ss}en}, {Jevardat de Fombelle}, {Jordan}, {Krone-Martins}, {Lanzafame}, {L{\"o}ffler}, {Marchal}, {Marrese}, {Moitinho}, {Muinonen}, {Osborne}, {Pancino}, {Pauwels}, {Recio-Blanco}, {Reyl{\'e}}, {Riello}, {Rimoldini}, {Roegiers}, {Rybizki}, {Sarro}, {Siopis}, {Smith}, {Sozzetti}, {Utrilla}, {van Leeuwen}, {Abbas}, {{\'A}brah{\'a}m}, {Abreu Aramburu}, {Aerts}, {Aguado}, {Ajaj}, {Aldea-Montero}, {Altavilla}, {{\'A}lvarez}, {Alves}, {Anders}, {Anderson}, {Anglada Varela}, {Antoja}, {Baines}, {Baker}, {Balaguer-N{\'u}{\~n}ez}, {Balbinot}, {Balog}, {Barache}, {Barbato}, {Barros}, {Barstow}, {Bartolom{\'e}}, {Bassilana}, {Bauchet}, {Becciani}, {Bellazzini}, {Berihuete}, {Bernet}, {Bertone}, {Bianchi}, {Binnenfeld}, {Blanco-Cuaresma}, {Blazere}, {Boch}, {Bombrun}, {Bossini}, {Bouquillon}, {Bragaglia}, {Bramante}, {Breedt},
  {Bressan}, {Brouillet}, {Brugaletta}, {Bucciarelli}, {Burlacu}, {Butkevich}, {Buzzi}, {Caffau}, {Cancelliere}, {Cantat-Gaudin}, {Carballo}, {Carlucci}, {Carnerero}, {Carrasco}, {Casamiquela}, {Castellani}, {Castro-Ginard}, {Chaoul}, {Charlot}, {Chemin}, {Chiaramida}, {Chiavassa}, {Chornay}, {Comoretto}, {Contursi}, {Cooper}, {Cornez}, {Cowell}, {Crifo}, {Cropper}, {Crosta}, {Crowley}, {Dafonte}, {Dapergolas}, {David}, {David}, {de Laverny}, {De Luise}, {De March}, {De Ridder}, {de Souza}, {de Torres}, {del Peloso}, {del Pozo}, {Delbo}, {Delgado}, {Delisle}, {Demouchy}, {Dharmawardena}, {Di Matteo}, {Diakite}, {Diener}, {Distefano}, {Dolding}, {Edvardsson}, {Enke}, {Fabre}, {Fabrizio}, {Faigler}, {Fedorets}, {Fernique}, {Fienga}, {Figueras}, {Fournier}, {Fouron}, {Fragkoudi}, {Gai}, {Garcia-Gutierrez}, {Garcia-Reinaldos}, {Garc{\'\i}a-Torres}, {Garofalo}, {Gavel}, {Gavras}, {Gerlach}, {Geyer}, {Giacobbe}, {Gilmore}, {Girona}, {Giuffrida}, {Gomel}, {Gomez}, {Gonz{\'a}lez-N{\'u}{\~n}ez},
  {Gonz{\'a}lez-Santamar{\'\i}a}, {Gonz{\'a}lez-Vidal}, {Granvik}, {Guillout}, {Guiraud}, {Guti{\'e}rrez-S{\'a}nchez}, {Guy}, {Hatzidimitriou}, {Hauser}, {Haywood}, {Helmer}, {Helmi}, {Sarmiento}, {Hidalgo}, {Hilger}, {H{\l}adczuk}, {Hobbs}, {Holland}, {Huckle}, {Jardine}, {Jasniewicz}, {Jean-Antoine Piccolo}, {Jim{\'e}nez-Arranz}, {Jorissen}, {Juaristi Campillo}, {Julbe}, {Karbevska}, {Kervella}, {Khanna}, {Kontizas}, {Kordopatis}, {Korn}, {K{\'o}sp{\'a}l}, {Kostrzewa-Rutkowska}, {Kruszy{\'n}ska}, {Kun}, {Laizeau}, {Lambert}, {Lanza}, {Lasne}, {Le Campion}, {Lebreton}, {Lebzelter}, {Leccia}, {Leclerc}, {Lecoeur-Taibi}, {Liao}, {Licata}, {Lindstr{\o}m}, {Lister}, {Livanou}, {Lobel}, {Lorca}, {Loup}, {Madrero Pardo}, {Magdaleno Romeo}, {Managau}, {Mann}, {Manteiga}, {Marchant}, {Marconi}, {Marcos}, {Marcos Santos}, {Mar{\'\i}n Pina}, {Marinoni}, {Marocco}, {Marshall}, {Polo}, {Mart{\'\i}n-Fleitas}, {Marton}, {Mary}, {Masip}, {Massari}, {Mastrobuono-Battisti}, {Mazeh}, {McMillan}, {Messina}, {Michalik},
  {Millar}, {Mints}, {Molina}, {Molinaro}, {Moln{\'a}r}, {Monari}, {Mongui{\'o}}, {Montegriffo}, {Montero}, {Mor}, {Mora}, {Morbidelli}, {Morel}, {Morris}, {Muraveva}, {Murphy}, {Musella}, {Nagy}, {Noval}, {Oca{\~n}a}, {Ogden}, {Ordenovic}, {Osinde}, {Pagani}, {Pagano}, {Palaversa}, {Palicio}, {Pallas-Quintela}, {Panahi}, {Payne-Wardenaar}, {Pe{\~n}alosa Esteller}, {Penttil{\"a}}, {Pichon}, {Piersimoni}, {Pineau}, {Plachy}, {Plum}, {Poggio}, {Pr{\v{s}}a}, {Pulone}, {Racero}, {Ragaini}, {Rainer}, {Raiteri}, {Rambaux}, {Ramos}, {Ramos-Lerate}, {Re Fiorentin}, {Regibo}, {Richards}, {Rios Diaz}, {Ripepi}, {Riva}, {Rix}, {Rixon}, {Robichon}, {Robin}, {Robin}, {Roelens}, {Rogues}, {Rohrbasser}, {Romero-G{\'o}mez}, {Rowell}, {Royer}, {Ruz Mieres}, {Rybicki}, {Sadowski}, {S{\'a}ez N{\'u}{\~n}ez}, {Sagrist{\`a} Sell{\'e}s}, {Sahlmann}, {Salguero}, {Samaras}, {Sanchez Gimenez}, {Sanna}, {Santove{\~n}a}, {Sarasso}, {Schultheis}, {Sciacca}, {Segol}, {Segovia}, {S{\'e}gransan}, {Semeux}, {Shahaf}, {Siddiqui}, {Siebert},
  {Siltala}, {Silvelo}, {Slezak}, {Slezak}, {Smart}, {Snaith}, {Solano}, {Solitro}, {Souami}, {Souchay}, {Spagna}, {Spina}, {Spoto}, {Steele}, {Steidelm{\"u}ller}, {Stephenson}, {S{\"u}veges}, {Surdej}, {Szabados}, {Szegedi-Elek}, {Taris}, {Taylo}, {Teixeira}, {Tolomei}, {Tonello}, {Torra}, {Torra}, {Torralba Elipe}, {Trabucchi}, {Tsounis}, {Turon}, {Ulla}, {Unger}, {Vaillant}, {van Dillen}, {van Reeven}, {Vanel}, {Vecchiato}, {Viala}, {Vicente}, {Voutsinas}, {Weiler}, {Wevers}, {Wyrzykowski}, {Yoldas}, {Yvard}, {Zhao}, {Zorec}, {Zucker}, \& {Zwitter}}]{GaiaDR3}
{Gaia Collaboration}, {Vallenari}, A., {Brown}, A.~G.~A., {$et~al$.} 2023, \aap, 674, A1

\bibitem[{Glushkova {$et~al$.}(2010)Glushkova, Koposov, Zolotukhin, Beletsky, Vlasov, \& Leonova}]{Glushkova2010}
Glushkova, E., Koposov, S., Zolotukhin, I.~Y., {$et~al$.} 2010, Astronomy Letters, 36, 75

\bibitem[{Hao {$et~al$.}(2022)Hao, Xu, Wu, Lin, Liu, \& Li}]{Hao2022}
Hao, C., Xu, Y., Wu, Z., {$et~al$.} 2022, Astronomy \& Astrophysics, 660, A4

\bibitem[{Hao {$et~al$.}(2021)Hao, Xu, Hou, Bian, Li, Wu, He, Li, \& Liu}]{Hao2021}
Hao, C., Xu, Y., Hou, L., {$et~al$.} 2021, Astronomy \& Astrophysics, 652, A102

\bibitem[{{Haroon} {$et~al$.}(2025{\natexlab{a}}){Haroon}, {Elsanhoury}, {Elkholy}, {naby Saad}, \& {{\c{C}}{\i}nar}}]{Haroon_2025}
{Haroon}, A., {Elsanhoury}, W., {Elkholy}, E., {naby Saad}, A., \& {{\c{C}}{\i}nar}, D.~C. 2025{\natexlab{a}}, Physica Scripta, 100, 055006

\bibitem[{Haroon {$et~al$.}(2014)Haroon, Ismail, \& Alnagahy}]{Haroon2014}
Haroon, A., Ismail, H., \& Alnagahy, F. 2014, Astrophysics and Space Science, 352, 665

\bibitem[{Haroon {$et~al$.}(2017)Haroon, Ismaill, \& Elsanhoury}]{Haroon2017}
Haroon, A., Ismaill, H., \& Elsanhoury, W. 2017, Astrophysics, 60, 173

\bibitem[{{Haroon} {$et~al$.}(2025{\natexlab{b}}){Haroon}, {Elsanhoury}, {Elkholy}, {Saad}, \& {{\c{C}}{\i}nar}}]{Haroon2025}
{Haroon}, A.~A., {Elsanhoury}, W.~H., {Elkholy}, E.~A., {Saad}, A.~S., \& {{\c{C}}{\i}nar}, D.~C. 2025{\natexlab{b}}, \physscr, 100, 055006

\bibitem[{{Huang} {$et~al$.}(2021){Huang}, {Yuan}, {Beers}, \& {Zhang}}]{Huang2021}
{Huang}, Y., {Yuan}, H., {Beers}, T.~C., \& {Zhang}, H. 2021, \apjl, 910, L5

\bibitem[{{Hunt} \& {Reffert}(2024)}]{Hunt2024}
{Hunt}, E.~L., \& {Reffert}, S. 2024, \aap, 686, A42

\bibitem[{{Juri{\'c}} {$et~al$.}(2008){Juri{\'c}}, {Ivezi{\'c}}, {Brooks}, {Lupton}, {Schlegel}, {Finkbeiner}, {Padmanabhan}, {Bond}, {Sesar}, {Rockosi}, {Knapp}, {Gunn}, {Sumi}, {Schneider}, {Barentine}, {Brewington}, {Brinkmann}, {Fukugita}, {Harvanek}, {Kleinman}, {Krzesinski}, {Long}, {Neilsen}, {Nitta}, {Snedden}, \& {York}}]{Juric2008}
{Juri{\'c}}, M., {Ivezi{\'c}}, {\v{Z}}., {Brooks}, A., {$et~al$.} 2008, The Astrophysical Journal, 673, 864

\bibitem[{{Karata{\c{s}}} {$et~al$.}(2023){Karata{\c{s}}}, {{\c{C}}akmak}, {Akkaya Oralhan}, {Bonatto}, {Michel}, \& {Netopil}}]{Karatas2023}
{Karata{\c{s}}}, Y., {{\c{C}}akmak}, H., {Akkaya Oralhan}, {\.I}., {$et~al$.} 2023, \mnras, 521, 2408

\bibitem[{Kharchenko {$et~al$.}(2016)Kharchenko, Piskunov, Schilbach, R{\"o}ser, \& Scholz}]{kharchenko2016}
Kharchenko, N., Piskunov, A., Schilbach, E., R{\"o}ser, S., \& Scholz, R.-D. 2016, Astronomy \& Astrophysics, 585, A101

\bibitem[{King(1962)}]{King1962}
King, I. 1962, The Astronomical Journal, 67, 471

\bibitem[{King(1966)}]{King1966}
King, I.~R. 1966, The Astronomical Journal, 71, 64

\bibitem[{Koester \& Reimers(1996)}]{Koester1996}
Koester, D., \& Reimers, D. 1996, Astronomy and Astrophysics, v. 313, p. 810-814, 313, 810

\bibitem[{Koposov {$et~al$.}(2008)Koposov, Glushkova, \& Zolotukhin}]{koposov2008}
Koposov, S., Glushkova, E., \& Zolotukhin, I.~Y. 2008, Astronomy \& Astrophysics, 486, 771

\bibitem[{{Krone-Martins} \& {Moitinho}(2014)}]{KroneMartins2014}
{Krone-Martins}, A., \& {Moitinho}, A. 2014, \aap, 561, A57

\bibitem[{{Kroupa}(2001)}]{Kroupa2001}
{Kroupa}, P. 2001, \mnras, 322, 231

\bibitem[{{K{\"u}pper} {$et~al$.}(2008){K{\"u}pper}, {MacLeod}, \& {Heggie}}]{2008MNRAS.387.1248K}
{K{\"u}pper}, A. H.~W., {MacLeod}, A., \& {Heggie}, D.~C. 2008, \mnras, 387, 1248

\bibitem[{Lada \& Lada(2003)}]{LadaandLada2003}
Lada, C.~J., \& Lada, E.~A. 2003, Annual Review of Astronomy and Astrophysics, 41, 57

\bibitem[{{Lindegren} {$et~al$.}(2021){Lindegren}, {Bastian}, {Biermann}, {Bombrun}, {de Torres}, {Gerlach}, {Geyer}, {Hern{\'a}ndez}, {Hilger}, {Hobbs}, {Klioner}, {Lammers}, {McMillan}, {Ramos-Lerate}, {Steidelm{\"u}ller}, {Stephenson}, \& {van Leeuwen}}]{Lindegren2021}
{Lindegren}, L., {Bastian}, U., {Biermann}, M., {$et~al$.} 2021, \aap, 649, A4

\bibitem[{{Liu} {$et~al$.}(2023){Liu}, {Xu}, {Hao}, {Bian}, {Lin}, {Li}, \& {Li}}]{Liu2023}
{Liu}, D., {Xu}, Y., {Hao}, C., {$et~al$.} 2023, \apjs, 268, 46

\bibitem[{Magrini {$et~al$.}(2009)Magrini, Sestito, Randich, \& Galli}]{Magrini2009}
Magrini, L., Sestito, P., Randich, S., \& Galli, D. 2009, Astronomy \& Astrophysics, 494, 95

\bibitem[{Maurya {$et~al$.}(2021)Maurya, Joshi, Elsanhoury, \& Sharma}]{Maurya2021}
Maurya, J., Joshi, Y., Elsanhoury, W., \& Sharma, S. 2021, The Astronomical Journal, 162, 64

\bibitem[{{Maurya} {$et~al$.}(2023){Maurya}, {Joshi}, {Samal}, {Rawat}, \& {Gour}}]{Maurya2023}
{Maurya}, J., {Joshi}, Y.~C., {Samal}, M.~R., {Rawat}, V., \& {Gour}, A.~S. 2023, Journal of Astrophysics and Astronomy, 44, 71

\bibitem[{Melchior(1958)}]{Melchior1958}
Melchior, P. 1958, Ciel et Terre, Vol. 74, p. 290, 74, 290

\bibitem[{Monteiro {$et~al$.}(2021)Monteiro, Barros, Dias, \& L{\'e}pine}]{Monteiro2021}
Monteiro, H., Barros, D.~A., Dias, W.~S., \& L{\'e}pine, J.~R. 2021, Frontiers in Astronomy and Space Sciences, 8, 656474

\bibitem[{Monteiro {$et~al$.}(2020)Monteiro, Dias, Moitinho, Cantat-Gaudin, L{\'e}pine, Carraro, \& Paunzen}]{Monteiro2020}
Monteiro, H., Dias, W., Moitinho, A., {$et~al$.} 2020, Monthly Notices of the Royal Astronomical Society, 499, 1874

\bibitem[{{Netopil} {$et~al$.}(2022){Netopil}, {Oralhan}, {{\c{C}}akmak}, {Michel}, \& {Karata{\c{s}}}}]{Netopil2022}
{Netopil}, M., {Oralhan}, {\.I}.~A., {{\c{C}}akmak}, H., {Michel}, R., \& {Karata{\c{s}}}, Y. 2022, \mnras, 509, 421

\bibitem[{Perren {$et~al$.}(2020)Perren, Giorgi, Moitinho, Carraro, Pera, \& Vazquez}]{Perren2020}
Perren, G.~I., Giorgi, E., Moitinho, A., {$et~al$.} 2020, Astronomy \& Astrophysics, 637, A95

\bibitem[{Perren {$et~al$.}(2015)Perren, Vazquez, \& Piatti}]{Perren2015}
Perren, G.~I., Vazquez, R.~A., \& Piatti, A.~E. 2015, Astronomy \& Astrophysics, 576, A6

\bibitem[{{Poggio} {$et~al$.}(2021){Poggio}, {Drimmel}, {Cantat-Gaudin}, {Ramos}, {Ripepi}, {Zari}, {Andrae}, {Blomme}, {Chemin}, {Clementini}, {Figueras}, {Fouesneau}, {Fr{\'e}mat}, {Lobel}, {Marshall}, {Muraveva}, \& {Romero-G{\'o}mez}}]{Poggio2021}
{Poggio}, E., {Drimmel}, R., {Cantat-Gaudin}, T., {$et~al$.} 2021, \aap, 651, A104

\bibitem[{Prusti {$et~al$.}(2016)Prusti, De~Bruijne, Brown, Vallenari, Babusiaux, Bailer-Jones, Bastian, Biermann, Evans, Eyer, {$et~al$.}}]{Prusti2016}
Prusti, T., De~Bruijne, J., Brown, A.~G., {$et~al$.} 2016, Astronomy \& astrophysics, 595, A1

\bibitem[{Salpeter(1955)}]{Salpeter1955}
Salpeter, E.~E. 1955, The Astrophysical Journal, 121, 161

\bibitem[{Sandage(1957)}]{Sandage1957}
Sandage, A. 1957, Astrophysical Journal, vol. 125, p. 435, 125, 435

\bibitem[{Sariya {$et~al$.}(2021)Sariya, Jiang, Sizova, Postnikova, Bisht, Chupina, Vereshchagin, Yadav, Rangwal, \& Tutukov}]{Sariya2021}
Sariya, D.~P., Jiang, G., Sizova, M., {$et~al$.} 2021, The Astronomical Journal, 161, 101

\bibitem[{Sim {$et~al$.}(2019)Sim, Lee, Ann, \& Kim}]{Sim2019}
Sim, G., Lee, S.~H., Ann, H.~B., \& Kim, S. 2019, arXiv preprint arXiv:1907.06872

\bibitem[{Skrutskie {$et~al$.}(2006)Skrutskie, Cutri, Stiening, Weinberg, Schneider, Carpenter, Beichman, Capps, Chester, Elias, {$et~al$.}}]{Skrutskie2006}
Skrutskie, M., Cutri, R., Stiening, R., {$et~al$.} 2006, The Astronomical Journal, 131, 1163

\bibitem[{{Soubiran} {$et~al$.}(2018){Soubiran}, {Cantat-Gaudin}, {Romero-G{\'o}mez}, {Casamiquela}, {Jordi}, {Vallenari}, {Antoja}, {Balaguer-N{\'u}{\~n}ez}, {Bossini}, {Bragaglia}, {Carrera}, {Castro-Ginard}, {Figueras}, {Heiter}, {Katz}, {Krone-Martins}, {Le Campion}, {Moitinho}, \& {Sordo}}]{Soubiran18}
{Soubiran}, C., {Cantat-Gaudin}, T., {Romero-G{\'o}mez}, M., {$et~al$.} 2018, \aap, 619, A155

\bibitem[{{Spina} {$et~al$.}(2021){Spina}, {Ting}, {De Silva}, {Frankel}, {Sharma}, {Cantat-Gaudin}, {Joyce}, {Stello}, {Karakas}, {Asplund}, {Nordlander}, {Casagrande}, {D'Orazi}, {Casey}, {Cottrell}, {Tepper-Garc{\'\i}a}, {Baratella}, {Kos}, {{\v{C}}otar}, {Bland-Hawthorn}, {Buder}, {Freeman}, {Hayden}, {Lewis}, {Lin}, {Lind}, {Martell}, {Schlesinger}, {Simpson}, {Zucker}, \& {Zwitter}}]{2021MNRAS.503.3279S}
{Spina}, L., {Ting}, Y.~S., {De Silva}, G.~M., {$et~al$.} 2021, \mnras, 503, 3279

\bibitem[{Spitzer~Jr \& Hart(1971)}]{Spitzer1971}
Spitzer~Jr, L., \& Hart, M.~H. 1971, Astrophysical Journal, vol. 166, p. 483, 166, 483

\bibitem[{{Ta\c sdemir} \& {Yontan}(2023)}]{Tasdemir2023}
{Ta\c sdemir}, S., \& {Yontan}, T. 2023, Physics and Astronomy Reports, 1, 1

\bibitem[{{Ta{\c{s}}demir} \& {{\c{C}}{\i}nar}(2025)}]{Tasdemir2025a}
{Ta{\c{s}}demir}, S., \& {{\c{C}}{\i}nar}, D.~C. 2025, arXiv e-prints, arXiv:2501.17235

\bibitem[{{Ta{\c{s}}demir} {$et~al$.}(2025){Ta{\c{s}}demir}, {{\c{C}}{\i}nar}, {Canbay}, {Taştan}, {Elsanhoury}, \& {Haroon}}]{Tasdemir2025b}
{Ta{\c{s}}demir}, S., {{\c{C}}{\i}nar}, D.~C., {Canbay}, R., {$et~al$.} 2025, arXiv e-prints, arXiv:2503.19015

\bibitem[{Tarricq {$et~al$.}(2022)Tarricq, Soubiran, Casamiquela, Castro-Ginard, Olivares, Miret-Roig, \& Galli}]{Tarricq2022}
Tarricq, Y., Soubiran, C., Casamiquela, L., {$et~al$.} 2022, Astronomy \& Astrophysics, 659, A59

\bibitem[{{{\v{S}}ablevi{\v{c}}i{\={u}}t{\.{e}}} {$et~al$.}(2006){{\v{S}}ablevi{\v{c}}i{\={u}}t{\.{e}}}, {Vansevi{\v{c}}ius}, {Kodaira}, {Narbutis}, {Stonkut{\.{e}}}, \& {Brid{\v{z}}ius}}]{2006BaltA..15..547S}
{{\v{S}}ablevi{\v{c}}i{\={u}}t{\.{e}}}, I., {Vansevi{\v{c}}ius}, V., {Kodaira}, K., {$et~al$.} 2006, Baltic Astronomy, 15, 547

\bibitem[{{Wang} {$et~al$.}(2022){Wang}, {Yuan}, \& {Huang}}]{Wang2022}
{Wang}, C., {Yuan}, H., \& {Huang}, Y. 2022, \aj, 163, 149

\bibitem[{Yadav {$et~al$.}(2014)Yadav, Leonova, Sagar, \& Glushkova}]{Yadav2014}
Yadav, R., Leonova, S., Sagar, R., \& Glushkova, E. 2014, Journal of Astrophysics and Astronomy, 35, 143

\bibitem[{{Yontan} {$et~al$.}(2022){Yontan}, {{\c{C}}akmak}, {Bilir}, {Banks}, {Ra{\'u}l}, {Canbay}, {Ko{\c{c}}}, {Ta{\c{s}}demir}, {Er{\c{c}}ay}, {Tan{\i}k Ozt{\"u}rk}, \& {Dursun}}]{Yontan2022}
{Yontan}, T., {{\c{C}}akmak}, T., {Bilir}, S., {$et~al$.} 2022, The Revista Mexicana de Astronomía y Astrofísica, 58, 333

\bibitem[{{Yucel} {$et~al$.}(2024){Yucel}, {Canbay}, \& {Bakis}}]{Yucel2024}
{Yucel}, G., {Canbay}, R., \& {Bakis}, V. 2024, Physics and Astronomy Reports, 2, 18

\bibitem[{Zhong {$et~al$.}(2019)Zhong, Chen, Kouwenhoven, Li, Shao, \& Hou}]{Zhong2019}
Zhong, J., Chen, L., Kouwenhoven, M., {$et~al$.} 2019, Astronomy \& Astrophysics, 624, A34

\bibitem[{{Zinn}(2021)}]{Zinn2021}
{Zinn}, J.~C. 2021, \aj, 161, 214

\end{thebibliography}

\end{document}